\documentstyle[preprint,epsfig,aps]{revtex}

\title{ 
{
\begin{flushright}
{\normalsize 
TUHEP-99-06\\
PDK-743\\}
\end{flushright}
\bf Search for Nucleon Decay into\\
 Lepton $+$ K$^0$ Final States Using Soudan 2}
      }
      
\author{
D.~Wall$^{5,a}$, 
W.W.M.~Allison$^3$, G.J.~Alner$^4$, D.S.~Ayres$^1$,
W.L.~Barrett$^6$,\\ 
C.~Bode$^2$, P.M.~Border$^2$, C.B.~Brooks$^3$, J.H.~Cobb$^3$, R.~Cotton$^4$,
H.~Courant$^2$,\\
D.M.~Demuth$^2$, T.H.~Fields$^1$, H.R.~Gallagher$^3$, C.~Garcia-Garcia$^4$,\\ 
M.C.~Goodman$^1$, R.~Gran$^2$, T.~Joffe-Minor$^1$, T.~Kafka$^5$,
S.M.S.~Kasahara$^2$,\\ 
W.~Leeson$^1$, P.J.~Litchfield$^4$, N.P.~Longley$^2$, W.A.~Mann$^5$,
M.L.~Marshak$^2$,\\
R.H.~Milburn$^5$, W.H.~Miller$^2$, L.~Mualem$^2$, A.~Napier$^5$,
W.P.~Oliver$^5$,\\
G.F.~Pearce$^4$, E.A.~Peterson$^2$, D.A.~Petyt$^4$, L.E.~Price$^1$,
K.~Ruddick$^2$,\\ 
M.~Sanchez$^5$, J.~Schneps$^5$, M.H.~Schub$^2$, R.~Seidlein$^1$,
A.~Stassinakis$^3$,\\
H.~Tom$^5$, J.L.~Thron$^1$, G.~Villaume$^2$, S.P.~Wakely$^2$, N.~West$^3$\\
\vspace{.15in}
$^{1}${\it Argonne National Laboratory, Argonne, IL 60439}\\
$^{2}${\it University of Minnesota, Minneapolis, MN 55455}\\
$^{3}${\it Department of Physics, University of Oxford, Oxford OX1 3RH, UK}\\
$^{4}${\it Rutherford Appleton Laboratory, Chilton, Didcot, Oxfordshire
 OX11 0QX, UK}\\
$^{5}${\it Tufts University, Medford, MA 02155}\\
$^{6}${\it Western Washington University, Bellingham, WA 98225}\\
$^a${Now at Sapient Corporation, One Memorial Drive, Cambridge, MA 02142}\\}

\begin{document}
\draft

\maketitle 
\thispagestyle{empty}

\begin{abstract}

A search for nucleon decay into two--body final states containing K$^0$
mesons has been conducted using the 963 metric ton Soudan 2 iron tracking
calorimeter. The topologies, ionizations, and kinematics of contained events
recorded in a 5.52 kiloton-year total exposure
(4.41 kton-year fiducial volume exposure) are examined for
compatibility with nucleon decays in an iron medium. For proton decay into the
fully visible final states $\mu^+$K$^0_s$ and e$^+$K$^0_s$, zero
and one event candidates are observed respectively. The lifetime lower limits
($\tau /B$) thus implied are $1.5 \times 10^{32}$ years and
$1.2 \times 10^{32}$ years, respectively. Lifetime lower limits are also
reported for proton decay into $\ell^+$K$^0_l$, and for neutron decay into
$\nu$K$^0_s$. 

\end{abstract}

\pacs{}

\section{Introduction}

\subsection{SUSY predictions for nucleon decay}

Supersymmetry (SUSY) proposes the existence, presumably at high mass scales, of
a new fermion or boson partner for each boson or fermion encompassed by the
Standard Model. At the expense of doubling the number of elementary particles,
SUSY Grand Unified Theories (GUTs) provide a natural solution to the hierarchy
problem, allow the extrapolations of the running coupling constants (which have
now been precisely determined at low energies and electroweak unification
scales) to converge at a high energy scale, make accurate predictions of the
Weinberg angle, and incorporate gravitation when locally gauged
\cite{susy:taylor_review,susy:kane_How,susy:superunification,susy:lopez_report}.

\begin{figure}[htb]
\centerline{\epsfig{figure=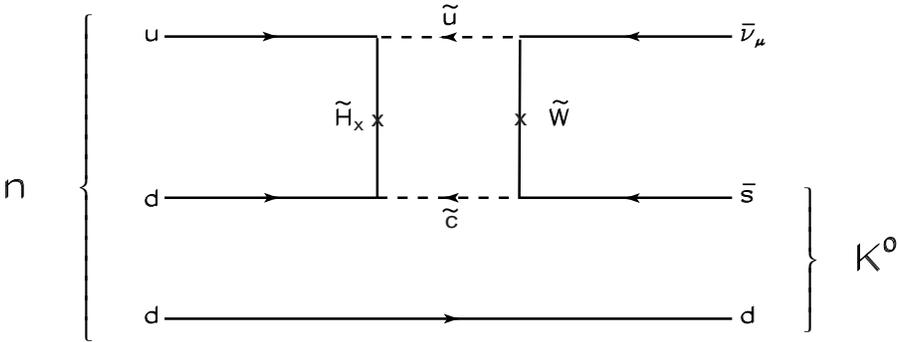,width=6in}}
\caption
{A nucleon decay amplitude in SUSY GUTs.}
\label{fig:susyfeyndg}
\end{figure}

Supersymmetric GUTs, such as SUSY SU(5), permit the same processes for nucleon
decay as do their non-supersymmetric counterparts; however, the larger mass
requirements for the GUT--scale lepto-quark bosons generally preclude the
possibility of observing nucleon decay with lifetime divided by branching ratio
$\tau$/B of less than $\sim 10^{33}$ years. Interestingly, other processes
involving supersymmetric particle loops arise that can mediate nucleon decay.
An example is shown in Fig.~\ref{fig:susyfeyndg} wherein the decay is suppressed
by the product of the color triplet Higgsino and Wino masses rather than by the
supersymmetric version of the X and Y lepto-quark bosons
\cite{theory:nath_arno_photino,theory:langacker_grand_unific}. Nucleon decay
diagrams of this type give integrals that vanish unless the transitions
involve intergenerational mixing.
Consequently, final states containing strange
mesons are to be expected, with lifetimes predicted to be $\sim 10^{31} -
10^{32}$ years \cite{theory:Hisano_and_Murayama,theory:lucas_and_rabi}.


Table I lists some proposed SUSY theories along with their
predicted decay modes of leading branching ratios and their lifetime
predictions. The expectation that two--body decays yielding a
positive--strangeness $K$ meson are predominant, is seen to be a common theme.

\subsection{Previous experimental searches for $K^0$ modes}

Experimental searches for baryon instability in general and for nucleon decay
into two-body modes containing K-mesons in particular, have been carried out
for more than two decades. Among the early searches which included strangeness
modes were the planar iron tracking calorimeter experiments of KGF
\cite{kgf:predpdk,kgf:predpdkncim}, NUSEX \cite{nusx:fcevts}, and Soudan 1
\cite{soud1:pdksrch}. Lifetime lower limits and also observation of a few
nucleon decay candidate events, were reported. Among candidate sightings was a
three-prong event recorded by NUSEX having kinematics compatible with

   \begin{equation}
{\rm p} \rightarrow \mu^+ {\rm K}^0 {\textstyle ,~~} {\rm K}^0_s
 \rightarrow \pi^+ \pi^-.
   \label{eq:threeProng}
   \end{equation}
 
With more exposure however, this candidate was found to be consistent with
background arising from interactions of atmospheric neutrinos within the NUSEX
calorimeter.

Lifetime lower limits for proton decay (\ref{eq:threeProng}) (but including 
the K$^0_s \rightarrow \pi^0 \pi^0$ mode) and for neutron decay

   \begin{equation}
{\rm n} \rightarrow \nu {\rm K}^0 {\textstyle,~~} {\rm K}^0_s
 \rightarrow \pi^0 \pi^0.
   \label{eq:nuKZero}
   \end{equation}
\noindent
were reported by the IMB water Cherenkov experiment in 1984 \cite{imb:lK}. These
limits were improved upon and the scope of search extended to other K-meson
modes in subsequent investigations by IMB
\cite{imb:23m,imb:haines,imb:multitrack}. Lifetime
limits were also obtained with the HPW water Cherenkov detector
\cite{hpw:nukmuon}. More recently, nucleon decay limits for many modes 
have been published, including the K$^0$ modes considered
in the present work, based upon the 7.6 kton-year exposure recorded by
IMB-3\cite{imb:mcgrew}.

Nucleon decays involving K$^0$ and K$^+$ mesons were searched for using the iron
tracking calorimeter experiment of the Frejus collaboration; lifetime lower
limits are reported in Refs.~\cite{frej:numes,frej:chlep,frej:b-lviol}. Among
the highest of published lifetime lower limits for lepton plus kaon two-body
modes are the
1989 results by the Kamiokande water Cherenkov experiment \cite{kam:lepmes}
based upon a 3.76 kton year exposure of KAM-I and KAM-II.

A search for the proton decay mode p $\rightarrow \nu$K$^+$ in the Soudan 2
experiment has been reported in a previous publication \cite{soud:nukplus}.
A more stringent lifetime limit for this mode has recently been published
by Super-Kamiokande \cite{SK_PRL83}. In this work we report our search
in Soudan 2 for nucleon decay into two--body modes involving
K$^0$ mesons, namely proton decay into $\mu^+$K$^0$ and e$^+$K$^0$ with
K$^0 \rightarrow$ K$^0_s$ or K$^0_l$, and neutron decay into $\nu$K$^0_s$.
Soudan 2 has the capability to observe these leading SUSY decays if indeed
their lifetimes are within the ranges indicated in Table I.

\section{Detector and Event Samples}

\label{sec:two}

\subsection{Tracking calorimeter and active shield}

Soudan 2 is a 963 metric ton iron tracking calorimeter which is
currently taking data. It is located at a depth of 2340 ft (2070 mwe) on the
27th level of the Soudan Underground Mine State Park in northern Minnesota.
Data--taking commenced in April 1989 when the detector's total mass was 275
tons. The modular design allowed data-taking to proceed
while additional 4.3-ton calorimeter modules were being installed.
The detector was completed in November 1993 at 963 tons.
The total (fiducial) exposure analyzed here is 5.52 (4.41)
kiloton--years (kty), obtained from data--taking through October 1998. 

Each of the 224 modules that comprise the calorimeter consists of
layered, corrugated iron sheets instrumented with drift tubes which are filled
with an Argon--CO$_{2}$ gas mixture. Electrons liberated by incident ionizing
particles drift to the ends of the tubes under the influence of an electric
field sustained by a voltage gradient which is applied along the tubes. The
drifted charge is collected by vertical anode wires held at a high voltage,
while horizontal cathode pad strips register the image charges. Two of the
position coordinates of the ionized track segments are determined by
identifying the anode wire and cathode pad strips excited; the third
coordinate is determined from the drift--time of the liberated charge once
the event ``start time'' has been
determined \cite{soud:design_and_constr,soud:s2detector}.

Surrounding the tracking calorimeter but mounted on the cavern walls and well
separated from calorimeter surfaces, is the 1700 m$^2$ active shield array
comprised of 2 -- 3 layers of proportional tubes \cite{soud:shield}. The
shield enables us to identify events which are not fully contained within the
calorimeter. In particular, it provides tagging of background events initiated
by cosmic ray muons, as will be described in sections \ref{sec:twob} and
\ref{sec:twod} below.

In searches for nucleon decay into multiprong modes, the Soudan 2 iron tracking
calorimeter of honeycomb lattice geometry offers event imaging capabilities
heretofore not achieved by the planar iron calorimeter experiments or by the
water Cherenkov experiments.
For the multiprong events of this work, Soudan's fine--grained tracking
enables event vertex locations to be established to within 0 to 3 cm in the
anode or cathode coordinate (drift-time coordinate) relative to the
honeycomb stack (to the rest of the event).
This spatial resolution enables
discrimination, based upon proximity to the event vertex, between a prompt
e$^{\pm}$ shower and a photon initiated shower. In Soudan 2, ionizing
particles having non-relativistic as well as relativistic momenta are imaged
with $dE/dx$ sampling.
Consequently, final state topologies are more completely ascertained, and proton
tracks can be distinguished from charged pion and muon tracks.
These capabilities are valuable for rejection of background.

\subsection{Event samples}
\label{sec:twob}

Events used in this analysis are required to be fully contained within a
fiducial volume, defined to be the 770-metric-ton portion of the calorimeter
which is everywhere more than 20 cm from all outer surfaces. There are four
distinct samples which are either simulated (Monte Carlo events) or
collected (data events) :

\begin{enumerate}

\item[{ }]~{\it i)} We use full detector nucleon decay Monte Carlo
simulations, including detector noise, to generate samples of the various
nucleon decay processes considered in this work. Hereafter we refer to these
events as {\it nucleon decay MC events}. Details concerning generation and
utilization of such events are given in Ref.~\cite{soud:nukplus}.

\item[{ }]~{\it ii)} A full detector neutrino Monte Carlo (MC) simulation,
which includes detector noise, is used to generate events
representing all charged-current and neutral-current reactions
which can be initiated by the flux of atmospheric neutrinos.
These {\it  neutrino MC events} are
injected into the data stream prior to physicist scanning and are subsequently
processed and evaluated using the same procedures as used with the data events.
The Monte Carlo program is based upon the flux predictions of Barr, Gaisser,
and Stanev \cite{flux:Bartol,flux:Barr} and is described in a
previous Soudan 2 publication \cite{soud:atmratio}.
Among the neutrino reactions simulated, care has been taken to include
K$^0$/$\overline {\rm K}^0$ channels; these arise at low rate from
$\Delta S = 0$ associated production and from $\Delta S = 1$
processes \cite{bubb:mann}. All of the atmospheric neutrino MC samples in this
paper correspond to a fiducial detector exposure of 24.0 kton-years.

\item[{ }]~{\it iii)} There are contained events for which the veto shield
registered coincident, double--layer hits. Such events usually originate with
inelastic interactions of cosmic ray muons within the cavern rock surrounding
the detector. The muon interactions eject charged and neutral particles out
of the rock and into the calorimeter; these particles subsequently interact to
make events in the detector. On rare occasions there
can be cosmic ray induced events which are unaccompanied by coincident hits in
the active shield; these arise either from shield inefficiency or from
instances wherein a neutron or photon with no accompanying charged particles
emerged from the cavern walls. Contained events of cosmic
ray origin are hereafter referred to as {\it ``rock events''}; these can be
{\it ``shield--tagged''} or otherwise of the {\it ``zero--shield--hit''}
variety.

\item[{ }]~{\it iv)} The {\it``gold events''} are data events for which the
cavern--liner active shield array registered no signal
during the event's allowed time window.
These events are mostly interactions of atmospheric neutrinos
\cite{soud:atmratio}.

\end{enumerate}

\subsection{Processing of events}

Each event included in a contained event sample for physics analysis has
passed through a standard data reduction chain. Care has been taken to ensure
that both data and Monte Carlo simulation events pass through the same
steps in the chain. For an event to be recorded, the
central detector requires at least seven contiguous pulse ``edges'' to occur
within a 72 $\mu$s window before a trigger is registered and the calorimeter
modules are read out. For MC events, the requirements of the
{\it hardware trigger} are imposed by a trigger simulation code. All events,
data and Monte Carlo, are then subjected to a {\it containment filter} code,
which rejects events for which part of the event extends outside the fiducial
volume. Taken together, the trigger and containment  filter requirements will
typically reduce a sample of MC nucleon decay events by 35\% to 65\%,
depending upon the decay mode being investigated.      

Events which survive the trigger and containment requirements, whether they be
data events (gold or rock) or Monte Carlo (neutrino reaction or nucleon decay),
are then subjected to two successive ``scanning passes'' carried out by
physicists \cite{soud:atmratio}. Each scanning pass involves three independent
scans. In the first pass, multiprong events are found with an overall
multiple-scan efficiency of $0.98^{+0.02}_{-0.04}$ \cite{scaneff}.
The first scan pass is designed to ensure {\it event quality}.
Here, residual uncontained events that have not been rejected by the software
filter are eliminated. Noise events, and events that have prongs ending on
detector cracks are also rejected at this stage. In the
second scanning pass, broad topology assignments are made (track or shower --
with or without a visible recoil proton, single proton and multiprong) and an
additional level of scrutiny with regard to containment is provided.
Multiprong events that survive both scanning passes are then subjected to
detailed {\it topology assignment}.

Once topologies have been assigned, events are reconstructed using software
tools available within our interactive graphics package. Tracks are fitted
using a routine that fits a polynomial curve through all hits selected and
returns the range, energy and initial direction vector. There is also a shower
processing routine which determines the energy and initial direction vector
for clusters of hits that are tagged as showers. The collection of
reconstructed tracks and showers along with mass hypotheses for the tracks is
then written into a data summary file, from which the derived quantities such
as event energy, invariant mass, net event momentum, etc., can be determined.
As a measure of event resolution, we calculate the difference between
Monte Carlo truth and the reconstructed final-state energy. The resulting
$\Delta E/E$ values for fully-visible final states are ($19 \pm 2$) \%
for multiprongs with prompt muons, and ($28 \pm 2$) \% for multiprongs with
prompt electrons. These values characterize reconstruction for nucleon
decay ($\ell^+$ + hadrons) events and for $\nu_\ell$ charged-current multiprongs
near threshold.

The fractions of event samples which remain after successive application
of each of the above-mentioned selections, averaged over the full volume of
the detector, are given in Table II 
for each nucleon decay mode of this study.
For each nucleon decay mode with a particular daughter process,
a sample of 493 MC events was generated.
For the K$^0_l$ modes, the portions of the MC samples which pass all cuts
except the kinematic ones are given by the top row, leftmost entries of
Tables III and IV.


\subsection{Cosmic ray induced background in ``gold" data}
\label{sec:twod}

       The gold data events analyzed in this work are multiprong events
which are mostly neutrino-induced, however a small number may
arise from rock events which are unaccompanied by hits in the veto shield
array.  The latter events are initiated by neutrons
emerging from the cavern rock which are incident upon the calorimeter.  In
contrast to neutrino interactions (or nucleon decay) which distribute 
uniformly throughout the calorimeter volume,  these events tend to occur
at relatively shallow depth into the detector.   

Our estimate of the number of zero-shield-hit rock events in multiprong data 
is based upon a multivariate discriminant analysis of the gold data,
neutrino MC, and shield-tagged rock samples\cite{Leeson:PDK684}.  For this 
analysis, discriminant functions characterizing each sample are constructed
using five event variables (of which several are highly correlated).  
These include: {\it i)} Penetration depth, measured along the event net
momentum, {\it ii)} vertex distance to the closest exterior surface, {\it iii)}
event visible energy, {\it iv)} zenith angle of the net momentum
vector, and {\it v)} ``inwardness", defined as the cosine between event 
$\vec{p}_{net}$ and the unit normal, pointing inward, of the closest exterior
surface of the calorimeter.  By fitting the discriminant variable 
distribution from gold data to a combination of neutrino MC and rock
discriminant distributions, the zero-shield-hit rock contribution is
ascertained. Out of the total number of multiprong rock events (with and
without shield hits), the fraction of events having shield hits is
$f = 0.94 \pm 0.04$. Among 144 gold data multiprongs, $16 \pm 11$ events 
may be cosmic-ray induced. Approximately 30\% of the zero-shield-hit rock rate
can be ascribed to inefficiency of the shield array.

      For each nucleon decay channel, we subject the shield-tagged rock
multiprongs to the same selection cuts which are applied to data
multiprongs.  The number of zero-shield-hit rock events which are
background for the channel is then calculated as the product $(1-f)/f$
times the number of shield-tagged rock events which pass the cuts.

\subsection{Neutrino background rates and atmospheric
$\nu_\mu$ oscillations}
\label{sec:IIE}

      In calculating event rates for neutrino-induced background in a
nucleon decay search, a consideration arises with the well-known
discrepancy between the observed versus predicted flavor ratio $\nu_\mu/\nu_e$
for atmospheric neutrinos.  In single track and single shower events,
Soudan 2 observes the ratio-of-ratios ($\nu_\mu/\nu_e$ observed/expected) to
be 0.64 $\pm$ 0.11 (stat.) $\pm$ 0.06 (syst.) \cite{soud:atmratioNEW}.
From the exposure analyzed for this work, we estimate
$128 \pm 14$ contained multiprong events to be neutrino events, where the
uncertainty is the quadrature sum of statistical error and an error from the
rock background subtraction. This can be compared to the number of
neutrino-induced multiprong events predicted by the null oscillation Monte
Carlo: $154 \pm 6$ events, where the uncertainty represents statistical error
from the atmospheric neutrino simulation. The difference may be
accounted for by atmospheric neutrino oscillations $\nu_\mu$ to $\nu_x$.
In any case, evidence for depletion of the atmospheric muon-neutrino flux is
sufficiently extensive \cite{SK:osc} that an accounting of the effect in
nucleon decay background rates is warranted.

The basis for our atmospheric neutrino background estimates is our
realistic MC simulation which uses null oscillation fluxes. The disappearance
of $\nu_\mu$-flavor neutrinos resulting from $\nu_\mu \rightarrow \nu_x$
oscillations, affects backgrounds initiated by $(\nu_\mu + \overline{\nu}_\mu)$
charged-current reactions; in fact, it reduces this background component.
In implementing our correction we further assume, as indicated by recent data,
that $\nu_x$ is an active neutrino which is not $\nu_e$ (i.e.
$\nu_x = \nu_\tau$) \cite{LP99}. Then, the predominant atmospheric
neutrino oscillation does not affect background arising from charged-current
reactions of $\nu_e$-flavor neutrinos, nor does it affect background from
neutral currents. Consequently, to correct for $\nu_\mu$-flavor disappearance,
the number of $(\nu_\mu + \overline{\nu}_\mu)$ charged-current background
events estimated from the atmospheric-neutrino MC for each nucleon decay
channel, has been scaled by the flavor ratio 0.64 $\pm$ 0.13.

As described below, the neutrino background is very low in 
the proton decay channels p $\rightarrow \mu^+ {\rm K}^0_s$,
K$^0_s \rightarrow \pi^+\pi^-$  and p $\rightarrow \mu^+ {\rm K}^0_l$. In
most other channels, e.g. p $\rightarrow \mu^+ {\rm K}^0_s$,
K$^0_s \rightarrow \pi^0\pi^0$, p $\rightarrow {\rm e}^+{\rm K}^0_s$,
p $\rightarrow {\rm e}^+ {\rm K}^0_l$, and
n $\rightarrow \nu {\rm K}^0_s$, K$^0_s \rightarrow \pi^0\pi^0$, the
neutrino background is either low or otherwise dominated by $\nu_e$ flavor
reactions. In fact, the only decay channel in this work for which
$\nu_\mu$-flavor disappearance
has a significant effect on the lifetime lower limit is neutron decay 
n $\rightarrow \nu {\rm K}^0_s$, K$^0_s \rightarrow \pi^+\pi^-$.
For this particular channel, backgrounds and corresponding
lifetime lower limits are presented both with and without correction for
$\nu_\mu$ oscillations.

\section{Characterization of Nucleon Decay Reactions}

For each nucleon decay mode, a Monte Carlo sample is created and processed as
described in Section~\ref{sec:two}; similarly handled are the gold data,
the contained shield-tagged rock events, and the atmospheric neutrino
MC events. These samples are used to determine the topological and kinematic
properties that differentiate each nucleon decay mode from other modes and
from the atmospheric neutrino and the rock event backgrounds.

\subsection{Event generation}

For each nucleon decay mode, five hundred events are typically generated
throughout the full detector volume. They are then overlaid onto pulser
trigger events which sample the detector background noise (from natural
radioactivity and cosmic rays) at regular intervals throughout the exposure. The
effects of Fermi motion within the iron and other nuclei which comprise the
detector are taken into account in simulation using the parametrization of
Ref.~\cite{theory:bodandrit}. For two--body nucleon decays in which the final
state momenta would otherwise be unique, Fermi motion smears the momenta and
thereby complicates final state identification.

For final states from neutrino interactions and also from nucleon decay
wherein pions are directly produced, intranuclear rescattering is treated in
our Monte Carlo using a phenomenological cascade model
\cite{mann:intranuke:leeson:pdk678}. Parameters of the model, which scale with
$A$, are set by requiring that threshold pion production observed in
$\nu_{\mu}$--deuteron ($A=2$) and $\nu_{\mu}$--neon ($A=20$) reactions
be reproduced \cite{merenyi:paper:thesis:merenyi}. For K$^0$ mesons however,
inelastic intranuclear rescattering is expected to be small due to the
absence of low--lying K$^0$N ($S = +1$) resonances and is not simulated.

\subsection{Final state kinematics}
\label{sec:threeb}
We examined each simulated nucleon decay mode for kinematic characteristics
which are suitable for selection of data candidates and for separation of
backgrounds. Two quantities that are generally useful are the invariant mass,
$M_{inv}$, and the magnitude of the net three--momentum,
$\left| \vec{p}_{net} \right|$, of reconstructed final state particles. For the
distribution in either variable, there is an expectation which can be
simulated using the nucleon decay MC events. For modes in which all final
state particles are visible, the invariant mass should fall in a range about
the nucleon mass (less 8 MeV binding energy) whose width is determined by
the resolution of the detector.

If some final state particles are not observable, useful information can still
be extracted from the invariant mass. For example, in the case of
n $\rightarrow \overline{\nu}$K$^0_s$ the mass will correspond to that of the
reconstructed K$^0_s$. The nominal 338 MeV/$c$ K$^0$ momentum from this two-body decay
will be boosted by the Fermi motion of the decaying neutron and smeared
further by detector resolution.


\subsection{{\bf Selection contour in the} $\bbox{M_{inv}}$ {\bf versus}
$\bbox{\left| \vec{p}_{net} \right|}$ {\bf plane}}
\label{sec:threec}


A scatter plot of invariant mass versus net event momentum can be created for
the reconstructed final states for each simulation. A region in this plane can
be chosen whose boundary defines a kinematical selection which can be
applied to the data and to the background samples. For our searches we consider
the application of such a cut to be the ``primary constraint'' and we have
formulated a quantitative parameterization of this two--variable constraint. We
observe that the final-state invariant mass exhibits a distribution which is
Gaussian to a good approximation.
Similarly, the final-state net momentum magnitude which results from the
folding of Fermi motion with the experimental resolution, distributes with a
shape which is very nearly Gaussian. Consequently, the density
distribution of points on the invariant mass versus momentum plane can be fitted
by a bi--variate Gaussian probability distribution function. In principle, when
the correlation coefficient is set to zero, the form of the fitting function
should be

   \begin{equation}
Z_{\alpha\beta}\left(x,y\right)=
\frac{1}{2\pi \sqrt{\sigma_{\alpha}\sigma_{\beta}}}
e^{-\frac{1}{2}\left[\left(\frac{x-\mu_
{\alpha}}{\sigma_{\alpha}}\right)^2 + \left(\frac{y-\mu_{\beta}}{\sigma_{\beta}}
\right)^2 \right]}
   \end{equation}
where $\sigma_{\alpha}$ and $\sigma_{\beta}$ are the widths and $\mu_{\alpha}$
and $\mu_{\beta}$ are the means of the two distributions. In practice however,
the conic section may have semimajor and semiminor axes that are not parallel
to the ordinate and abscissa of the graph. To allow such a rotation on the
plane, a new basis (which is a linear combination of the old variables) is
constructed with the new mean and widths defined on that basis. The form of
the fitting function then becomes
  \begin{equation}
    Z_{\alpha\beta}\left(x,y\right)=Ae^{-\frac{1}{2}\left[\left(\frac{ax + by
 -\mu_{\alpha}}
    {\sigma_{\alpha}}\right)^2 +
    \left(\frac{ay - bx - \mu_{\beta}}{\sigma_{\beta}}\right)^2 \right]}.
\label{Eq:3.2}
   \end{equation}
\noindent 
Note that the two arguments of the exponentials are formulated to be  mutually
perpendicular lines which, when fit, will be collinear with the major and
minor axes of the elliptical contours of the sections of the Gaussian
surface.

\begin{figure}[htb]
\centerline{\epsfig{figure=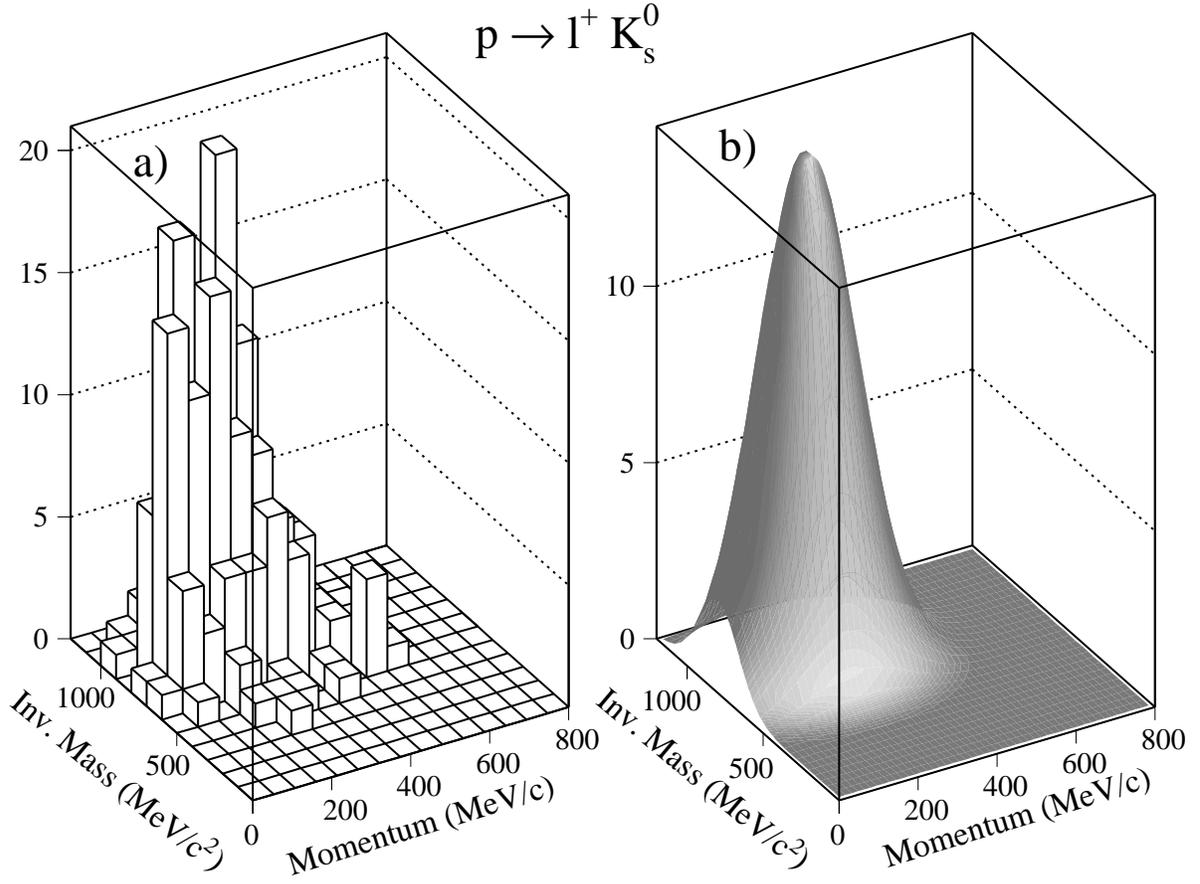}}
\caption{a) Simulation for decay mode
 p $\rightarrow \ell^+$K$^0_s$,
 displayed as a lego plot in $M_{inv}$ vs. $\left| \vec{p}_{net} \right|$
 space. b) The bi--variate Gaussian surface fitted to the distribution in
 Fig.~\ref{fig:legofit}a.}
\label{fig:legofit}
\end{figure}

\begin{figure}[htb]
\centerline{\epsfig{figure=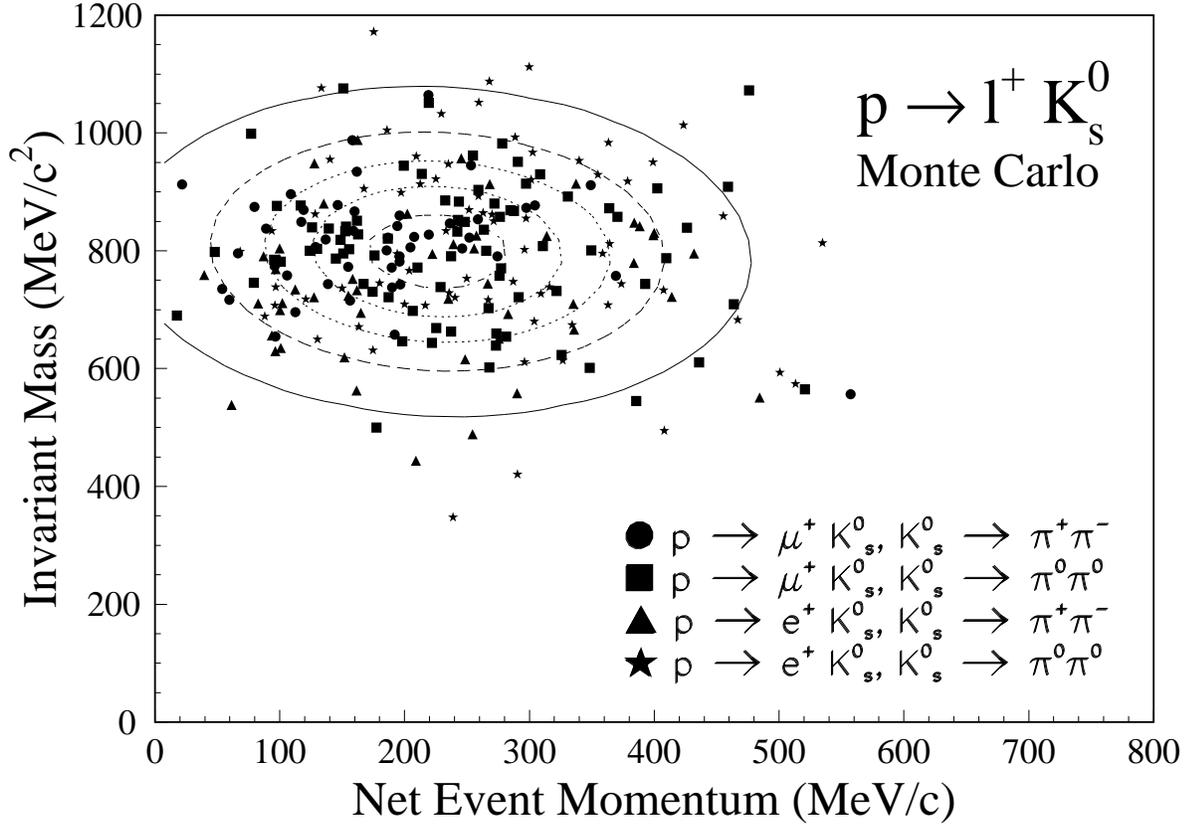,width=6.5in}}
\caption {
{The contours of the bi--variate Gaussian surface of
Fig.~\ref{fig:legofit} projected onto in the $M_{inv}$ vs.
$\left| \vec{p}_{net} \right|$ plane. The region bounded by the outermost
contour contains 90\% of reconstructed proton decay events and defines the
primary kinematic selection used to identify proton decay candidates.}
}
\label{fig:lpk0cont}
\end{figure}

For each mode in which this representation of a density distribution
is used, a lego plot for the simulation is
created (see Fig.~\ref{fig:legofit}a). The seven--parameter function above
is fitted to the simulation using a log-likelihood ${\chi}^2$ minimization
routine available in the MINUIT software package (see Fig.~\ref{fig:legofit}b
and Fig.~\ref{fig:lpk0cont}). The bi-variate Gaussian form
(Eq.~\ref{Eq:3.2}) was found to provide a good fit to the nucleon decay event
density distribution in every case, with $\chi^2$ per degree of freedom (dof)
values in the range 0.7 to 1.3.

      The projection of the bi-variate Gaussian surface of Fig. 2b onto 
the $M_{inv}$ versus $\left| \vec{p}_{net} \right|$ plane is shown in
Fig.~\ref{fig:lpk0cont}.  Here, the shape of the event distribution over the
plane is depicted using five nested, elliptical contour boundaries. Proceeding
outward from the innermost contour, the bounded regions contain respectively
10\%, 30\%, 50\%, 70\%, and 90\% of the event sample.  From the five regions
delineated, we choose the 90\%-of-sample contour - the outermost, solid curve
ellipse in Fig.~\ref{fig:lpk0cont} - for quantitative use in nucleon decay
searches of this study. In particular, for  p $\rightarrow \ell^+ {\rm K}^0_s$
searches we take this outermost contour to define our ``primary" kinematic
selection, namely:  A candidate proton decay event must have reconstructed
($M_{inv}$,  $\left| \vec{p}_{net} \right|$) values which
lie within the contour boundary.  Concerning the interior
contours (dashed ellipses of Fig.~\ref{fig:lpk0cont}), no quantitative
utilization will be made. However, by superposing event coordinates onto a
contour set such as shown in Fig.~\ref{fig:lpk0cont}, a degree of insight is
afforded as to whether an event sample as a whole exhibits the kinematics of
nucleon decay.  For this reason we display interior contours along with the
90\% contour in various kinematic diplots of this study.

\subsection{{\bf Other kinematic selection contours}}
\label{sec:threed}
The kinematic requirement on final-state invariant mass versus net momentum
combinations which the outermost contour of Fig.~\ref{fig:lpk0cont}
represents, is a constraint which nucleon decay will satisfy but which
background processes need not respect. The background rejection thus afforded
can be augmented using contours constructed  with other kinematic variables.

From trials with Monte Carlo event samples we find that, with procedures
to be described below, the final-state lepton prong can be identified
with near certainty (in final states e$^+$($\pi^+\pi^-$) and
$\mu^+$($\pi^0\pi^0$)) or with useful efficiency (84\% for
$\mu^+$($\pi^+\pi^-$) and 82\% for e$^+$($\pi^0\pi^0$)).
With this identification in place, a distinction can be made between
the lepton and the K$^0_s$ system. (Unfortunately, the neutral track gap
produced by the moving K$^0_s$ is usually not large enough to be resolved
in the Soudan 2 calorimeter.) Then, the kinematics allowed for the K$^0_s$
subsystem of proton decay can usefully be specified by a contour boundary
in the plane of K$^0_s$ invariant mass versus K$^0_s$ momentum.
A further kinematic restriction on momentum-sharing between the lepton and
the K$^0_s$ can also be formulated (albeit mostly redundant with constraints
already mentioned) in terms of an allowed contour region in the lepton
momentum versus K$^0_s$ momentum plane.

One might expect that reconstruction of the charged pion tracks in
p $\rightarrow \ell^+$K$^0_s$ final state with K$^0_s \rightarrow \pi^+\pi^-$
in Monte Carlo samples would on average yield the K$^0$ invariant mass.
In the Soudan detector, however, scattering processes tend to reduce
the range of pion tracks. As a result, our reconstruction  of MC event
images  which is based upon track range, yields an invariant mass
distribution with a mean of 461 MeV/$c^2$. Also, the momentum distribution
of the reconstructed K$^0$ is shifted downward by about 50 MeV/$c$.

In our $\ell^+$K$^0_s$ and $\nu$K$^0_s$ searches described below,
supplementary requirements are invoked using the above-mentioned
additional search contours. Their construction proceeds similarly as with
the primary constraint contour of Sect.~\ref{sec:threec};
each contour represents the projection of a bivariate Gaussian fit to the
distribution of nucleon decay MC events in two kinematic variables.
A data event, in order to qualify as a nucleon decay candidate, must have
kinematic coordinates occuring within each contour projection, the boundary of
which, by construction, includes the coordinates of 90\% of the MC nucleon
decay sample.

\section{Nucleon decay search}

\label{sect:four}

\subsection{Search for $\bbox{{\rm p} \rightarrow \mu^+ {\rm K}^0_s}$ and 
$\bbox{{\rm p} \rightarrow {\rm e}^+ {\rm K}^0_s}$}

We now consider searches for two--body decay modes $\ell^+$K$^0_s$, for which
the entire final state is observable in Soudan 2. Although the standard SUSY
SU(5) GUT models usually predict lifetimes for $\ell^+$K$^0$ modes to be two
orders of magnitude longer than those for $\overline{\nu}$K modes
\cite{theory:murayama}, there are other GUT models in which a charged lepton
mode could be enhanced \cite{susy:carone,theory:babu_barr,babu_pati}.

\begin{figure}
\centerline{\epsfig{figure=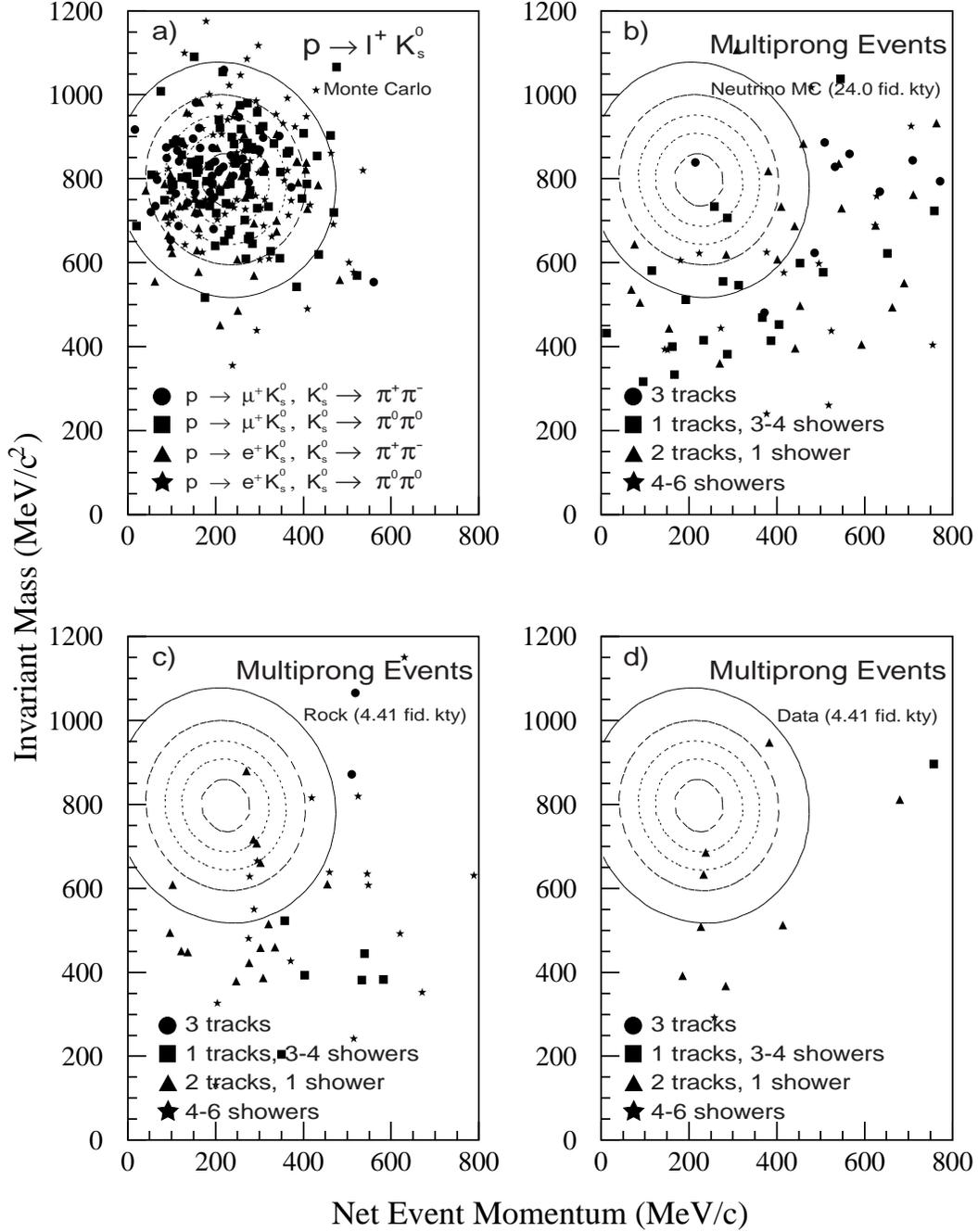,height=7in}}
\caption{For proton decay modes p $\rightarrow \ell^+$K$^0_s$,
the ``primary'' kinematic selection contour (outermost contour)
 together with event distributions,
in the $M_{inv}$ versus $\left| \vec{p}_{net} \right|$ plane. Distributions
show a) the proton decay simulations, b) atmospheric neutrino MC events,
c) rock events, and d) data events. The $\mu^+$K$^0_s$ (e$^+$K$^0_s$)
final states are depicted using solid circles and squares (triangles and
stars).}
\label{fig:lpk0con_main_quad}
\end{figure}

With $\ell^+$K$^0_s$ modes, 
the final state will be reconstructed with invariant mass
approximating the nucleon mass and with event net momentum within a range
compatible with Fermi motion in iron folded with the detector resolution. The
kinematically allowed region in the plane of final-state invariant mass versus
event net momentum, which is obtained from Soudan reconstructions
of Monte Carlo proton decay events, is rather
similar for all modes of this kind. As shown in Figure~\ref{fig:lpk0cont} and
also in Fig.~\ref{fig:lpk0con_main_quad}a, we have used simulations involving
four separate proton decay modes including K$^0_s$, to define a
``primary'' kinematic selection contour in the $M_{inv}$ versus $\left|
\vec{p}_{net} \right|$ plane. That is, for the proton decay searches involving
p $\rightarrow \mu^+$K$^0_s$ and p $\rightarrow$ e$^+$K$^0_s$, we
require all candidate events to have kinematics compatible with the outermost
contour depicted in Fig.~\ref{fig:lpk0con_main_quad}a. As indicated by
Figs.~\ref{fig:lpk0con_main_quad}b,~\ref{fig:lpk0con_main_quad}c and
~\ref{fig:lpk0con_main_quad}d, the primary contour alone provides a very
restrictive kinematic selection.

\subsubsection*{$\bbox{{\rm p} \rightarrow \mu^+ {\rm K}^0_s}$} 



\begin{figure}[htb]
\centerline{\epsfig{figure=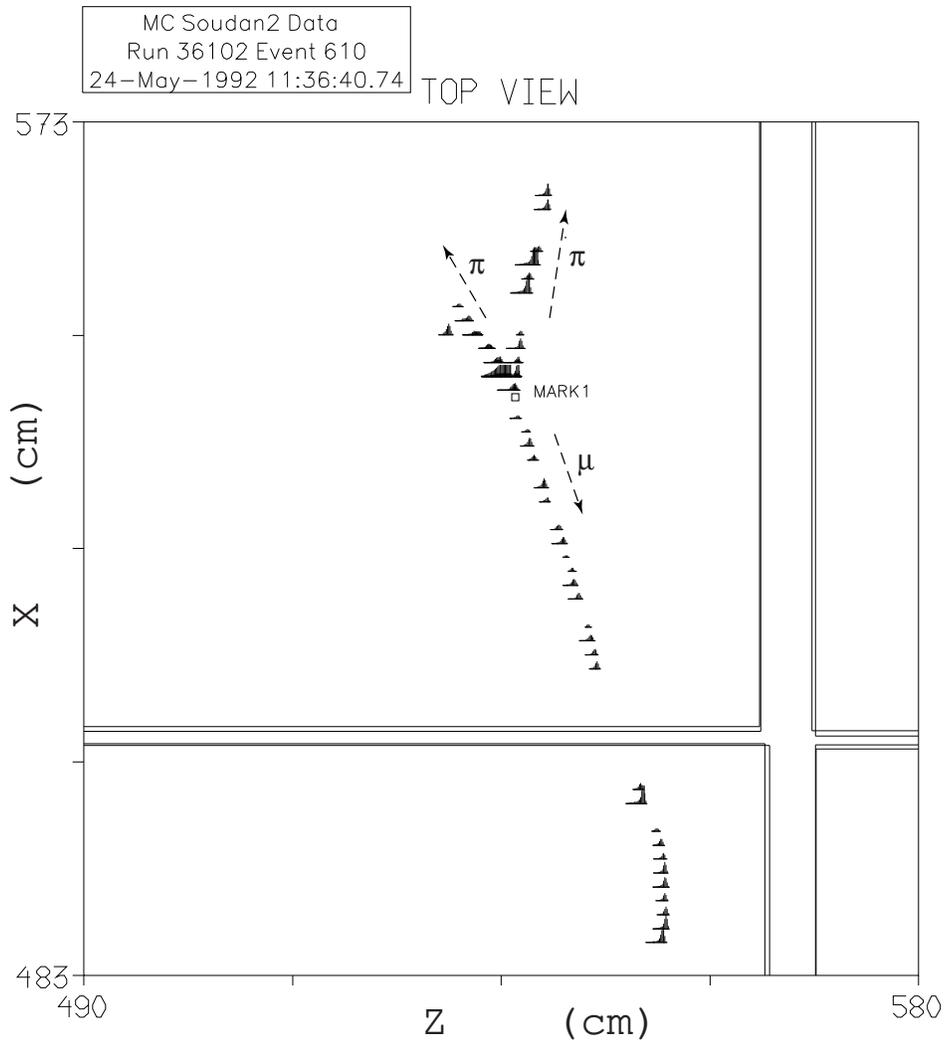,width=24.0cm}}
\caption{Monte Carlo event with full detector response for
proton decay p $\rightarrow \mu^+$K$^0_s$, K$^0_s \rightarrow \pi^+\pi^-$.
Here, the anode ($X$) versus drift time ($Z$)
projection has been selected from the three scanning views which display
anode-time, cathode-time, and anode-cathode images. The open square denoted
`MARK 1' depicts the reconstructed primary vertex.}
\label{fig:mc36102}
\end{figure}
\begin{figure}
\centerline{\epsfig{figure=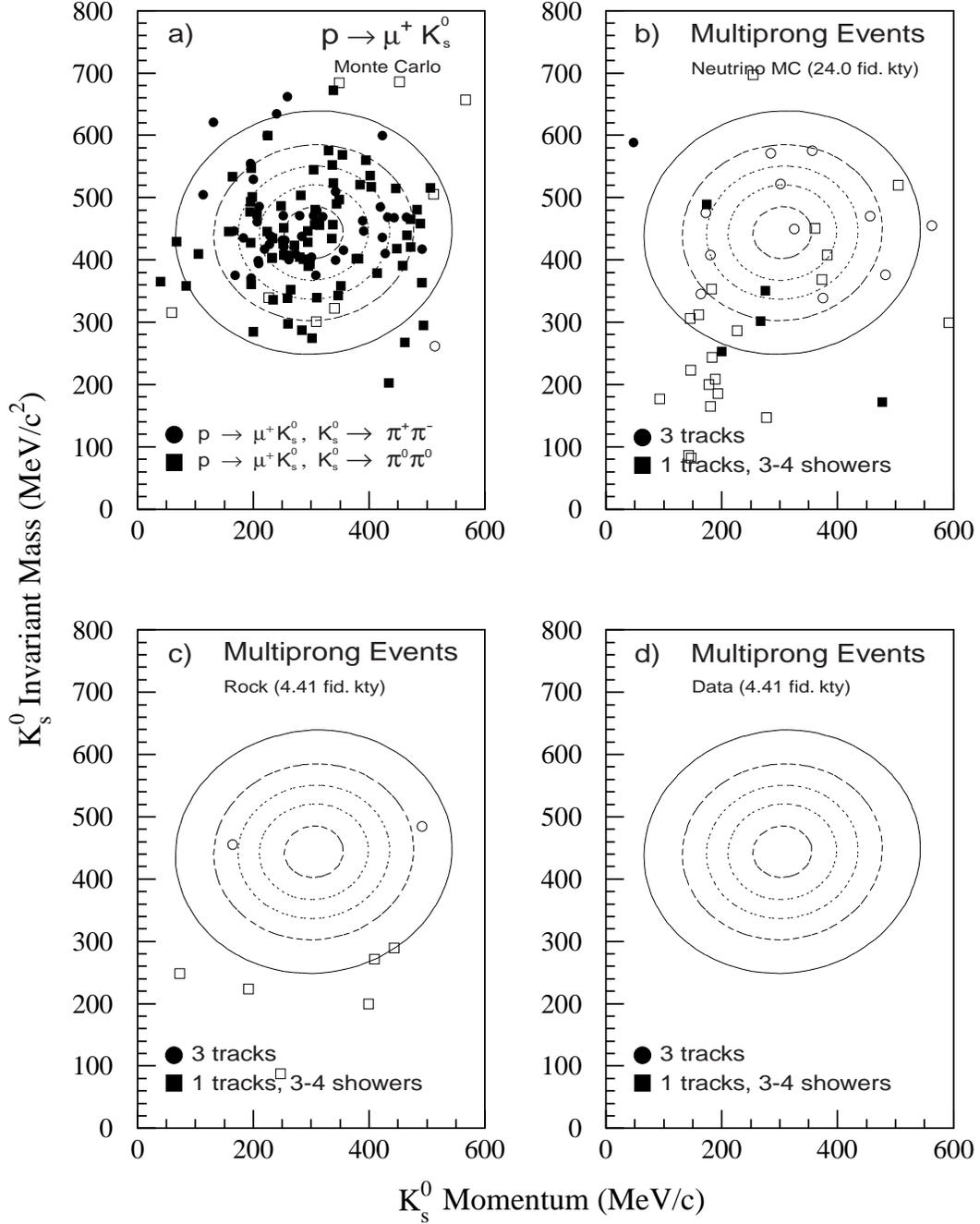,height=7in}}
\caption{For the K$^0_s$ subsystem of
 p $\rightarrow \mu^+$K$^0_s$ candidates, event distributions plus kinematic
selection contour in the plane of $M_{inv}$ versus
 $\left| \vec{p}_{{\rm K}^0} \right|$. Distributions
show a) proton decay simulations, b) atmospheric neutrino MC events, c) rock
events, and d) data events. Open symbol events fall outside the primary
contour of Fig.~\ref{fig:lpk0con_main_quad}.}
\label{fig:lpk0con_quad1}
\end{figure}

For this mode we examine two topologies: {\it i)} three tracks (from
K$^0_s \rightarrow \pi^+ \pi^-$) and {\it ii)} one track plus 3--4 showers (from
K$^0_s \rightarrow \pi^0 \pi^0$). In the first case, the track which
traverses the most calorimeter material and which does
not have a visible scatter is taken to be the $\mu^+$; the other two
tracks are tagged as pions. An illustrative Monte Carlo event of
p $\rightarrow \mu^+$K$^0_s$ with the three--track topology, is shown in
Fig.~\ref{fig:mc36102}. For K$^0_s \rightarrow \pi^0 \pi^0$
the lone track is assumed to be the muon, and the showers
are reconstructed as the $\pi^0 \pi^0$ system from the K$^0_s$. To make
optimal use of constraining variables, it is useful and also sufficient to
invoke here two distinct kinematic contours. In addition to requiring the
primary contour constraint on the final
state $M_{inv}$ versus $\left| \vec{p}_{net} \right|$, we require the
``K$^0$'' invariant mass versus momentum to fall within the contour of
Fig.~\ref{fig:lpk0con_quad1}. The
distributions of each of our four event samples (simulation, neutrino MC, rock
and data) with respect to the two different kinematic selection contours, for
both final states corresponding to p $\rightarrow \mu^+$K$^0_s$, are summarized
in Figures \ref{fig:lpk0con_main_quad} and \ref{fig:lpk0con_quad1}. In
Fig.~\ref{fig:lpk0con_quad1}, events that failed the preceeding
contour cut are shown using open symbols. A candidate event is required to
fall within both contours.
(Note that events in Fig.~\ref{fig:lpk0con_main_quad} can populate overflow
regions of Fig.~\ref{fig:lpk0con_quad1}, e.g. the track-plus-showers
event in Fig.~\ref{fig:lpk0con_main_quad}d is off-scale in
Fig.~\ref{fig:lpk0con_quad1}d.)

As a possible addition or alternative to the contour of
Fig.~\ref{fig:lpk0con_quad1}, event kinematics can be evaluated using a contour
in the plane of lepton momentum versus K$^0_s$ momentum. For our
$\mu^+$K$^0_s$ search, we find no improvement to be afforded. (For
e$^+$K$^0_s$ however, such a contour provides extra background
suppression and we utilize it in that search.)

 For the K$^0_s \rightarrow \pi^+ \pi^-$ case, the
contour cuts are 88\% efficient whereas for K$^0_s \rightarrow \pi^0 \pi^0$
the combined contour cut efficiency is 80\%. The application of these
kinematic cuts, together with the branching ratio, software and scanning
efficiencies, results in an overall proton-decay survival fraction of
$(16 \pm 3)$\% for the K$^0_s \rightarrow \pi^+ \pi^-$ case
and $(6 \pm 1)$\% for K$^0_s \rightarrow \pi^0 \pi^0$. No
event is observed to satisfy all kinematic constraints for either of the
$\pi \pi$ final states.

For the $\mu^+$K$^0_s$ final state with K$^0_s \rightarrow \pi^0 \pi^0$,
the estimated neutrino background of 0.6 events is mostly due to $\nu_e$
charged-current $\pi^+ \pi^0$ production. No neutrino MC events passed
the cuts for the $\mu^+ {\rm K}^0_s$ final state with
K$^0_s \rightarrow \pi^+ \pi^-$;
neutrino--induced backgrounds are expected to be low for this case
\cite{bubb:mann}.

From our observations of the K$^0_s \rightarrow \pi^+ \pi^-$ and $\pi^0 \pi^0$
decay modes, we set a proton decay lifetime
limit using the formula \cite{frej:numes,soud:nukplus,Wall:thesis}
   \begin{equation}
    \tau / {\textstyle B} ({\rm p} \rightarrow \mu^+ {\rm K}^0_s) > N_p
 \times T_f \times 
    \frac{\left[ \epsilon_1 \times {\textstyle B}{_1}(K) + 
           \epsilon_2 \times {\textstyle B}{_2}(K) \right]}
           {\mu_1 + \mu_2}.
   \label{eq:dualLimitCalc}
   \end{equation}
Here $N_{p(n)} = 2.87 (3.15) \times 10^{32}$ protons (neutrons) in a kiloton
of the Soudan 2 detector, $T_f = 5.52$ kiloton years is the full detector
exposure (4.41 kty is the fiducial exposure), and
$\epsilon_i \times {\textstyle B}{_i}(K)$ are the
selection efficiencies given in Table II. 
The $\mu_i$ are the constrained 90\% CL upper limits on the numbers of
observed events, and are found by solving the equation
   \begin{equation}
   0.10 = \frac{\sum_{n_1 = 0}^{n_{ev;1}}\sum_{n_2 = 0}^{n_{ev;2}}
                P(n_1, b_1 + \mu_1)P(n_2, b_2 + \mu_2)}
   {\sum_{n_1 = 0}^{n_{ev;1}}\sum_{n_2 = 0}^{n_{ev;2}}P(n_1, b_1)P(n_2, b_2)}
   \label{eq:mucalc}
   \end{equation}
with the constraint
   \begin{equation}
    \frac{\epsilon_1 \times {\textstyle B}{_1}(K)}{\mu_1} = 
    \frac{\epsilon_2 \times {\textstyle B}{_2}(K)}{\mu_2} =
\frac{\sum_{i=1}^2 \epsilon_i \times {\textstyle B}{_i}(K)}{\sum_{i=1}^2 \mu_i}
   \label{eq:constraints}
   \end{equation}
where $P(n,\mu)$ is the Poisson function, $e^{-\mu}\mu^n / n!$, and the $b_i$
are the estimated backgrounds. With no candidates found for either
$\mu^+$K$^0_s$ mode, we obtain $\mu_1 = \mu_2 = 2.31$. The combined
lower lifetime limit, at 90\% CL and with background subtraction, is then
$\tau / {\textstyle B} > 150 \times 10^{30}$~years.

Our limit for p $\rightarrow \mu^+ {\rm K}^0_s$ can be compared to the channel
limit previously obtained with the Frejus planar iron calorimeter:
$\tau/B > 64 \times 10^{30}$ years at 90\% CL \cite{frej:chlep}.
The water Cherenkov experiments Kamiokande and
IMB-3 have reported limits for p $\rightarrow \mu^+ {\rm K}^0$ which are
more stringent. Lifetime limits for $\mu^+ {\rm K}^0$ are summarized below,
following the discussion of p $\rightarrow \mu^+ {\rm K}^0_l$ searches.

\subsubsection*{$\bbox{{\rm p} \rightarrow {\rm e}^+{\rm K}^0_s}$} 

\begin{figure}
\centerline{\epsfig{figure=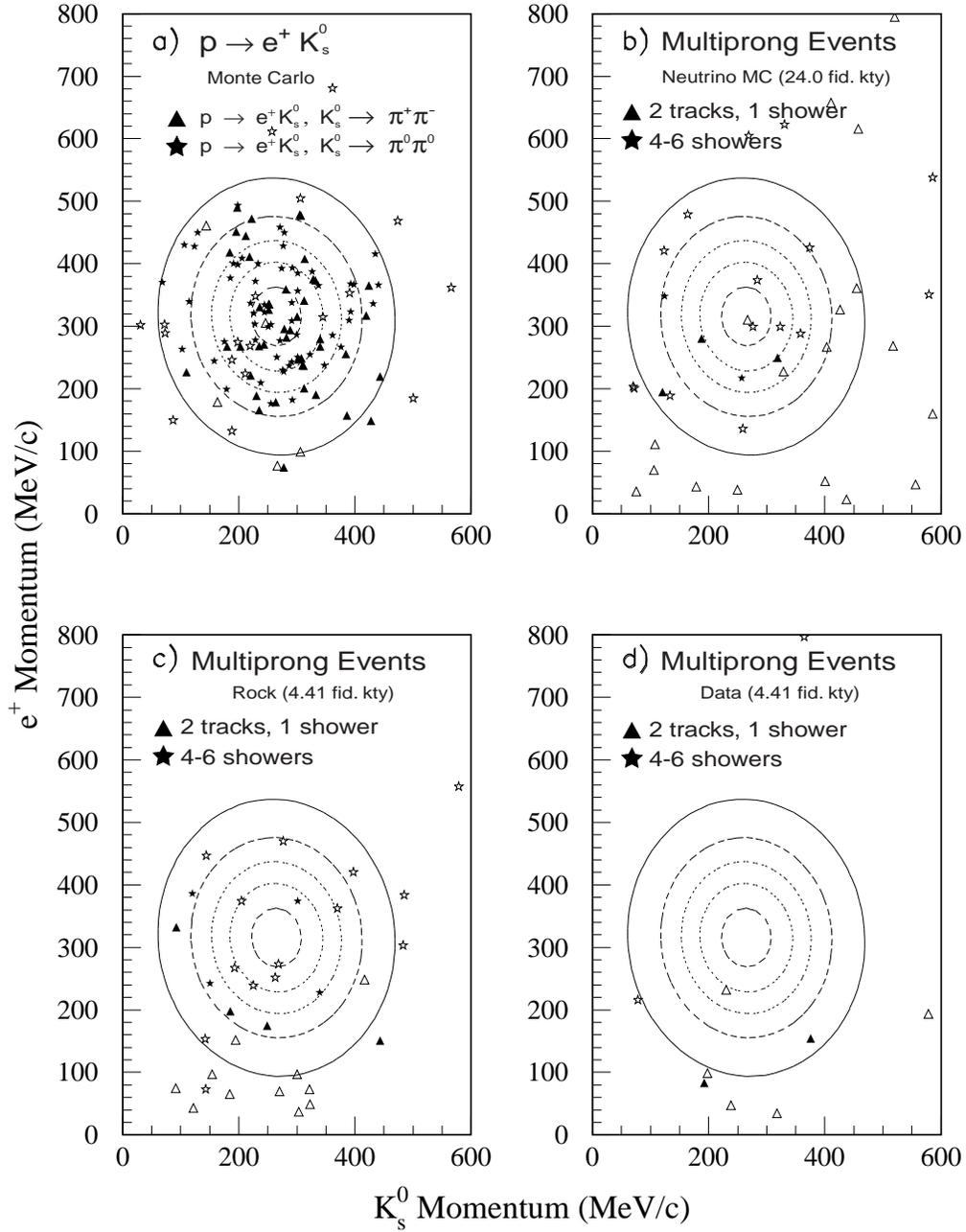,height=7in}}
\caption{The plane of e$^+$ momentum versus ``K$^0_s$''
momentum for p $\rightarrow$ e$^+$K$^0_s$ candidates, showing event
distributions plus kinematic selection contours. Distributions show a) proton
decay simulations, b) atmospheric neutrino MC events, c) rock events, and d)
data events. Open symbols denote events that fall outside of the primary
contour and/or the K$^0_s$ contour of the K$^0_s$ subsystem contour.}
\label{fig:lpk0con_quad4}
\end{figure}

We examine events which have the following topologies: {\it i)} one shower
plus two tracks (from K$^0_s \rightarrow \pi^+ \pi^-$), or
{\it ii)} 4 to 6 separate showers (from K$^0_s \rightarrow \pi^0 \pi^0$).
For p $\rightarrow$ e$^+$K$^0_s$, three kinematic constraint contours are
generated for each of the two daughter modes, and the data events and both
backgrounds are subjected to the cuts thus specified (see sections
\ref{sec:threec} and \ref{sec:threed}).
Figure~\ref{fig:lpk0con_main_quad} shows the primary contour ($M_{inv}$
vs. $\left| \vec{p}_{net} \right|$) for this case; Fig.~\ref{fig:lpk0con_quad4}
shows the final contour (e$^+$ momentum versus K$^0_s$ momentum) after the
K$^0$ mass versus K$^0$ momentum contour constraint has been applied
(the latter contour, not shown, introduces constraints as in
Fig~\ref{fig:lpk0con_quad1}.). In case {\it ii)} with 4--6 showers in the
final state, the prompt e$^+$ shower must be
distinguished from the $\gamma$ showers originating with decaying $\pi^0$s. In
scanning of candidate events, the shower prong most likely to be the positron
is chosen based upon the topology at the primary vertex. The momentum of the
shower chosen by the physicist scanner is subsequently compared to the
highest momentum among the other showers in the event.
If the physicists' choice is within 50 MeV/$c$ of being
the overall shower of highest momentum, it is accepted as the e$^+$, otherwise
the shower of highest momentum is chosen. In simulation, this algorithm
selects the true e$^+$ shower in 82\% of the events.
Again, events that have failed a previous cut are plotted using open symbols.
Events considered to be candidates are required to be inside of all three
contours.

In the case of the e$^+$K$^0_s$ final state with 
K$^0_s \rightarrow \pi^+ \pi^-$, the efficiency of all
three kinematic selections combined is 86\%. The overall efficiency times
branching ratio is then 15\%. The number of candidates observed
is one event whereas the total expected background is 0.74 events.
 Roughly two-thirds of the
expected background comes from $\nu_e$ charged-current single charged-pion
production events in which the proton is misidentified as another pion.

\begin{figure}[htb]
\centerline{\epsfig{figure=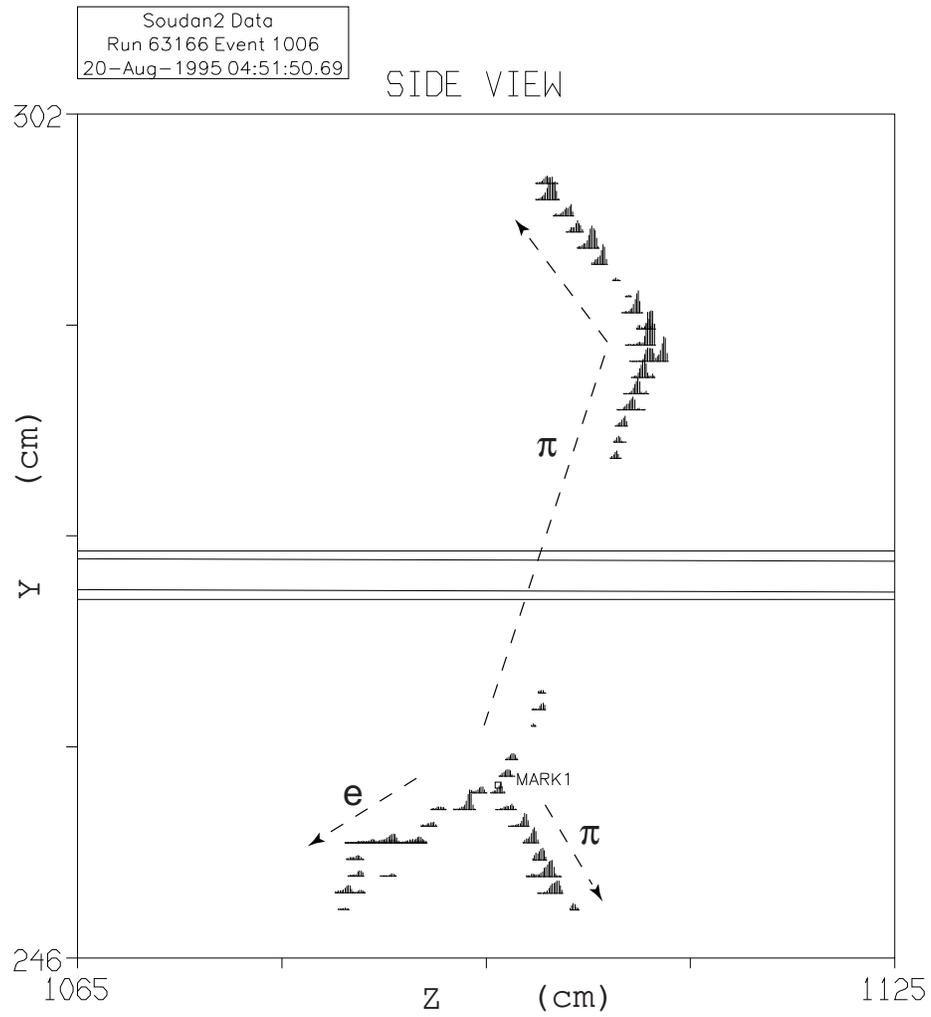,width=20.0cm}}
\caption{Data candidate for proton decay
 p $\rightarrow$ e$^+$K$^0_s$, K$^0_s \rightarrow \pi^+ \pi^-$,
shown in the cathode ($Y$) versus drift time ($Z$) projection.}
\label{fig:data63166}
\end{figure}

Fig.~\ref{fig:data63166} shows our single e$^+$K$^0_s$ candidate in the
most orthogonal
projection which for this event is the cathode ($Y$) versus drift time ($Z$)
projection. A ``prompt" electron shower and two tracks are seen to emerge
from the primary vertex (denoted MARK 1). The mean pulse height for either
track is too low for a proton assignment but is typical for a
pion (or muon). The longer pion track traverses an insensitive
module-to-module boundary region, giving rise to the apparent gap in the
track. The ionization pulse heights on this pion track
indicate that the vertex at higher $Y$ is a secondary scatter, not the 
primary vertex. These pulse heights are smaller (larger) before (after)
the proposed scatter point. The kinematics of this event is, however,
somewhat atypical for proton decay. In the plane of e$^+$ versus K$^0$
momentum, this event (depicted by the solid triangle in 
Fig.~\ref{fig:lpk0con_quad4}d)
lies within our search contour, but outside the contour
which contains 80\% of proton decays surviving our selections for this
mode.
It is of interest to consider the ``maximum" number of proton decays which
could appear as in Fig.~\ref{fig:data63166} and which are allowed in our
exposure, assuming the limits published by Kamiokande, Frejus, and IMB-3
\cite{kam:lepmes,frej:chlep,imb:mcgrew} to be correct.
Using the range of limit values reported for the  e$^+$K$^0_s$ mode
(see below) and the detection efficiency for this channel in the Soudan 2
detector (Table II, third row), 
 of order 0.8 to 3.2 occurrences
of proton decay resembling the Fig.~\ref{fig:data63166} candidate would be
compatible with previous limits. 

 For the e$^+$K$^0_s$ final state with K$^0_s \rightarrow \pi^0 \pi^0$,
75\% of all simulated proton decay events survive the three kinematic cuts
and the overall efficiency times branching ratio for this mode is
$(8 \pm 1)$\%. For this case no data events pass the cuts and 0.63 background
events are expected.

Combining the observations from the two separate decay sequences, we obtain
for ${\rm p} \rightarrow {\rm e}^+{\rm K}^0_s$ a lifetime lower limit at
90\% CL of $\tau / B > 120 \times 10^{30}$~years. 
A value of $\tau/B > 76 \times 10^{30}$ years was reported for this mode
 previously by Frejus \cite{frej:chlep}.
 Limits for e$^+$K$^0$ are summarized below, following the
discussion of p $\rightarrow {\rm e}^+ {\rm K}^0_l$ searches.


\subsection{Search for $\bbox{{\rm p} \rightarrow \mu^+ {\rm K}^0_l}$ and  
$\bbox{{\rm p} \rightarrow {\rm e}^+ {\rm K}^0_l}$}

For proton decay p $\rightarrow \ell^+$K$^0$, the  $K^0_s$ daughter decays
account for half of the possible final states; we now consider
p $\rightarrow \ell^+$K$^0_l$. In these decays the K$^0_l$ will interact
hadronically in the Soudan 2 medium before it has time to undergo $3 \pi$
decay. The event topologies
produced by the products of the K$^0_l$ interaction are quite varied but are
usually easily discernible. The basic topology searched for is that of a
single isolated lepton (either $\mu^+$ or e$^+$), with the appropriate
two--body decay momentum, separated from a secondary vertex produced by the
hadronic products of the K$^0_l$-nucleus interaction.
 Additional selections based on
the ``co-linearity'' of the event and the characteristic energy visible from
the K$^0_l$ interaction products are employed to provide discrimination from
atmospheric neutrino induced background events and also from inelastic
interactions of neutrons originating from cosmic ray collisions in the cavern
rock.

\subsubsection*{$\bbox{{\rm p} \rightarrow \mu^+ {\rm K}^0_l}$} 



\begin{figure}[htb]
\centerline{\epsfig{figure=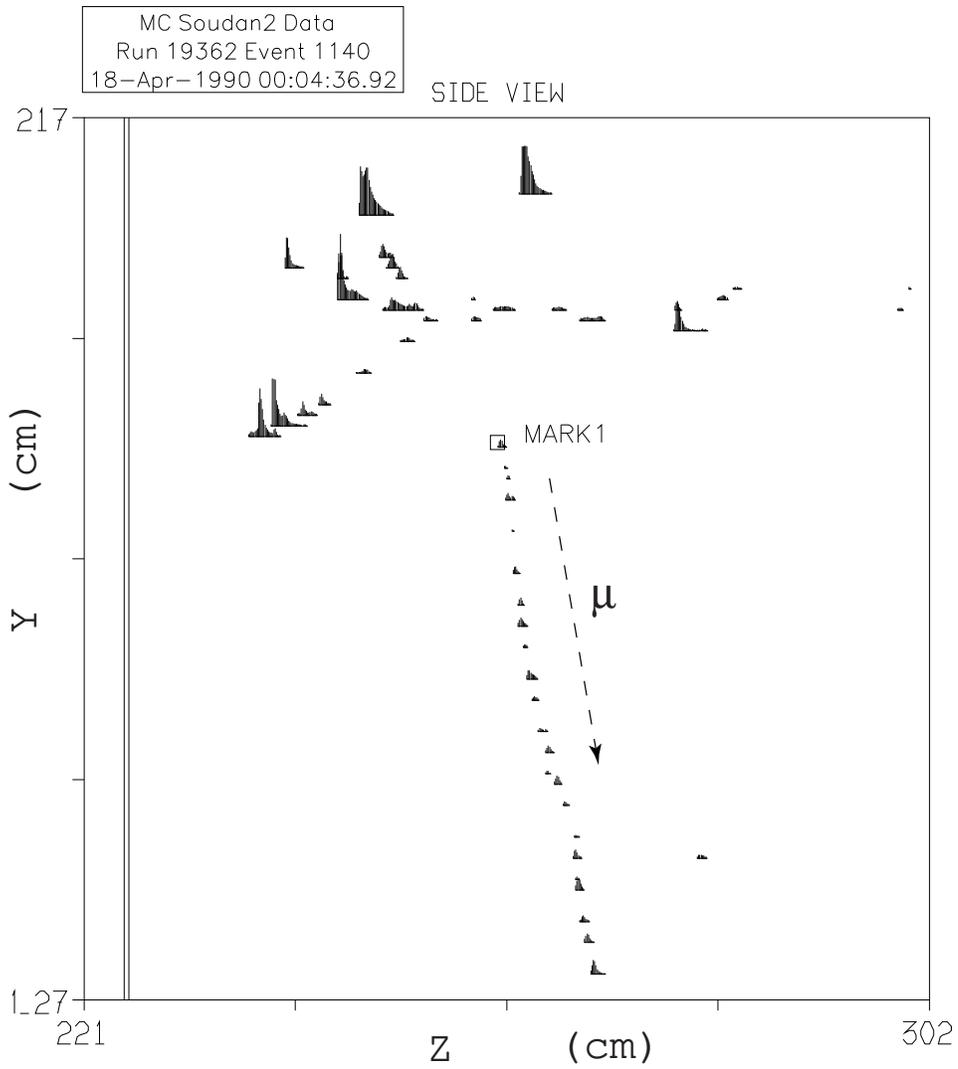,width=24.0cm}}
\caption{Monte Carlo event for p $\rightarrow \mu^+$K$^0_l$ with
 subsequent K$^0_l$ interaction in the calorimeter medium,
shown in the cathode versus drift time projection.}
\label{fig:mukl19362}
\end{figure}

We select events whose topology consists of an isolated muon track which appears
separated from a hadronic interaction vertex. A Monte Carlo event for this
process is shown in Fig.~\ref{fig:mukl19362}. This topology selection is
applied to each of the four samples of interest, namely the
p $\rightarrow \mu^+$K$^0_l$ full detector simulation, the neutrino reactions
of the atmospheric $\nu$ Monte Carlo, the shield-tagged rock events, and gold
data events. The sample populations thus obtained are given in the top row of
Table III; the neutrino background rate (column 3) has been corrected for
$\nu_\mu$-flavor depletion by $\nu_\mu \rightarrow \nu_x$ oscillations as
described in Section \ref{sec:IIE}.

 Kinematic cuts are then applied sequentially; their nature and order are
summarized in Table III, 
and the corresponding event populations which
survive each cut are listed for each of the four samples. We describe below each
of the kinematic cuts individually, in the order in which they are applied.

The isolated lepton is imaged as a single track with ionization compatible
with that of a muon mass assignment. The lone muon track is required to have a
length between 35 and 120 cm. We then require the gap between the event vertex,
identified by the start of the $\mu^+$ track, and the K$^0_l$ interaction to
be greater than 15 cm. The mean path gap for reconstructed proton-decay MC
events is 30 cm, however the
distribution is broad and extends beyond 100 cm. Since the secondary
interactions exhibit a wide range of topologies with no single topology
occuring very frequently, we estimate the visible energy of the hadronic final
states in a crude way by reconstructing entire final states as single
showers. That is, the visible energy is taken to be proportional to the total
number of gas tube crossings (``hits'') among the  associated tracks and
showers, with each tube hit requiring about 15 MeV of energy. This
procedure is carried out using the Soudan 2 shower processor; the processor
fits a direction vector to the interaction products as well as assigning an
interaction energy. In the events of the proton decay simulation, this
estimated K$^0_l$ interaction energy is observed to fall within the range
200 to 1400 MeV. This interval therefore represents our third kinematic
selection.

\begin{figure}
\centerline{\epsfig{figure=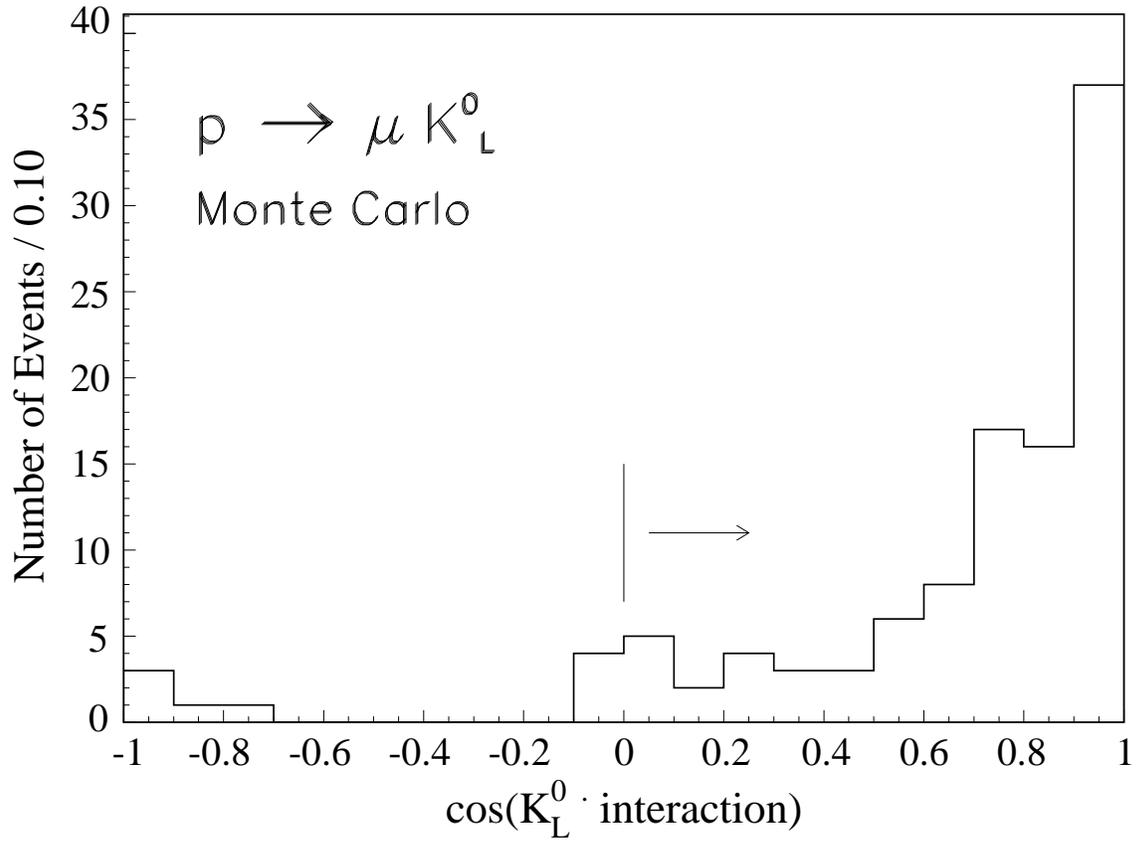,width=6.5in}}
\caption {
{Angle between the K$^0_l$ flight path and visible net momentum
of its subsequent hadronic interaction.}
  }
\label{fig:muklkintang}
\end{figure}

\begin{figure}
\centerline{\epsfig{figure=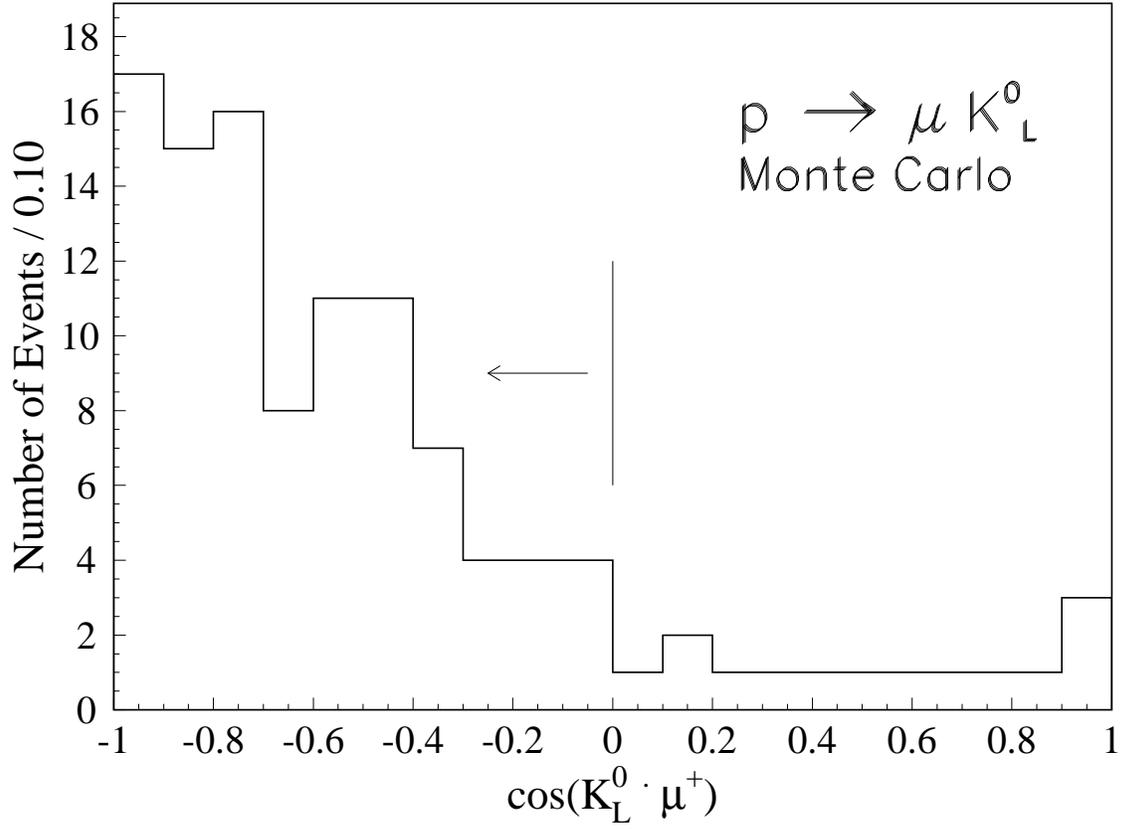,width=6.5in}}
\caption {
{Angle between the $\mu^+$ track and the K$^0_l$ flight path,
for p $\rightarrow \mu^+$K$^0_l$.}
  }
\label{fig:muklkmuang}
\end{figure}

It is to be expected that the K$^0_l$ interaction products will generally
travel in the original direction of the K$^0_l$. We then expect that the
vector from the event vertex to the K$^0_l$ interaction and the direction
vector returned by the shower processor from the K$^0_l$ interaction products,
point into the same hemisphere. Finally, since p $\rightarrow \mu^+$K$^0_l$ is
a two--body decay, the $\mu^+$ and the K$^0_l$ emerge back--to--back
in the rest frame of the parent proton. In the laboratory frame,
Fermi motion  can alter this back-to-back configuration. However,
in almost every event of the simulation, the reconstructed
$\mu^+$ direction and the K$^0_l$ path gap vector point in opposite
hemispheres. These angular cuts form the requirement that the
event exhibit a ``co--linearity'' such that the $\mu^+$, the K$^0_l$, and its
interaction products are all roughly aligned. The angular distributions for
the proton decay simulation are depicted in
Figures~\ref{fig:muklkintang}~and~\ref{fig:muklkmuang}.

The survival efficiency for proton decay to satisfy the kinematic cuts alone
is 53\%. Combining this efficiency with survival rates from the hardware
trigger simulation, the fiducial volume cut, plus scanning and topology cuts
we obtain an overall detection efficiency for p $\rightarrow \mu^+$K$^0_l$ of
$(12 \pm 1)$\%. Background from atmospheric neutrinos arises mostly from
$\nu_\mu$-flavor charged-current reactions. With correction for
$\nu_\mu$ oscillations, a rate of 0.2 background events
is estimated for the current exposure. A similar background rate is estimated
for rock events. The 90\% CL lower lifetime limit based on the K$^0_l$
component only is $\tau / B > 83 \times 10^{30}$~years.
The p $\rightarrow \mu^+ {\rm K}^0_l$ lifetime limit of
$44 \times 10^{30}$ years was reported previously by
Frejus \cite{frej:chlep}.

From our search results for $\mu^+ {\rm K}^0_s$ and $\mu^+ {\rm K}^0_l$,
we determine a lifetime lower limit for proton decay into $\mu^+ {\rm K}^0$
(with the K$^0 \rightarrow$ K$^0_{s,l}$ branching fractions included) in the
Soudan 2 experiment: $\tau/B > 120 \times 10^{30}$ years at 90\% CL.
Limits for $\mu^+ {\rm K}^0$ which are similarly prepared (90\% CL,
background subtracted), have been published previously by
Kamiokande \cite{kam:lepmes}, Frejus \cite{frej:chlep}, and
IMB-3 \cite{imb:mcgrew}; the limit values are
(120, 54, 120) $\times\ 10^{30}$ years respectively.

\subsubsection*{$\bbox{{\rm p} \rightarrow {\rm e}^+ {\rm K}^0_l}$} 


%
%
%
%
%
%
%
%
%
%

As in the case for p $\rightarrow \mu^+$K$^0_l$, we search for a single isolated
lepton prong which appears separated from a secondary multiprong vertex. We
first require the shower from the prompt positron to fall within an energy
range of 120 MeV to 500 MeV. The K$^0_l$ 15 cm path gap cut is then invoked,
followed by the two distinct angle cuts which require an overall co--linearity
of the shower and the gap and the hadronic interaction secondaries. The numbers
of background and candidate events which survive these selections are much
larger for this e$^+$K$^0_l$ search than for the $\mu^+$K$^0_l$ case, as
can be seen by comparing the populations tallied in columns two through four
of Tables III and IV.
This situation arises because multiprong events, not
infrequently, have gamma conversions which are remote from primary vertices
and thus mimic the topology of p $\rightarrow$ e$^+$K$^0_l$.

The interaction energy of the K$^0_l$ is again constructed by fitting a single
shower to the interaction products. 
For e$^+$K$^0_l$, the selection previously invoked
for the K$^0_l$ apparent energy in the $\mu^+$K$^0_l$ search is made more
restrictive; here we require 200 to 1100 MeV.
The cuts and their effects on the simulation, data and background samples are
summarized in Table IV. 

For our p $\rightarrow$ e$^+$K$^0_l$ search, the efficiency for survival through
the kinematic cuts is 50\% and the e$^+$K$^0_l$ detection efficiency is
$(11 \pm 1)$\%. From our null oscillation atmospheric neutrino sample, 43\% of
events originate with $\nu_\mu$-flavor charged-current reactions and receive 
correction for $\nu_\mu \rightarrow \nu_x$ oscillations. The total neutrino
background ($\nu_e$ and $\nu_\mu$ flavor, charged and neutral currents) is
2.6 events. Our estimate for
rock background without accompanying hits in the active shield is 0.8 events.
We find two gold data events which satisfy all of our e$^+$K$^0_l$
selections. The resulting 90\% CL lower lifetime limit for
p $\rightarrow$ e$^+$K$^0_l$ is then $\tau /B > 51 \times 10^{30}$~years.
The 90\% CL limit for p $\rightarrow e^+ {\rm K}^0_l$ previously reported
by Frejus is $\tau/B > 44 \times 10^{30}$ years \cite{frej:chlep}.

From the above results we determine a limit for proton decay into e$^+$K$^0$
with subsequent K$^0_s$ decay or K$^0_l$ interaction:
$\tau/B > 85 \times 10^{30}$ years at 90\% CL. Limits for e$^+$K$^0$
previously obtained by Kamiokande \cite{kam:lepmes}, Frejus \cite{frej:chlep}, and
IMB-3 \cite{imb:mcgrew}, are (150, 60, 31) $\times 10^{30}$ years.

\subsection{Search for $\bbox{{\rm n} \rightarrow \nu {\rm K}^0_s}$}

In SUSY GUTs, the mode n $\rightarrow \overline{\nu}$K$^0$ is similar to
p $\rightarrow  \overline{\nu}$K$^+$ in terms of its underlying amplitude (see
Fig.~\ref{fig:susyfeyndg}). However, predictions for the relative rates,
$\Gamma ({\rm n} \rightarrow \overline{\nu}_{\mu} {\rm K}^0) /
 \Gamma ({\rm p} \rightarrow \overline{\nu}_{\mu} {\rm K}^+)$ can vary
 significantly \cite{theory:lucas_and_rabi,theory:murayama}. It is possible
 that amplitude factors conspire to make n~$\rightarrow \overline{\nu}$K$^0$
 predominant, and so a thorough search is warranted.

For neutron decay into $\nu$K$^0_s$, K$^0_s \rightarrow \pi^+\pi^-$, events
consisting of two charged (non--proton) tracks are expected. For decays
leading to K$^0_s \rightarrow \pi^0\pi^0$, we expect four (or three) detectable
showers of modest energies to result from decays of the $\pi^0$s.

\subsubsection*{$\bbox{{\rm n} \rightarrow \nu {\rm K}^0_s}$, 
                $\bbox{{\rm K}^0_s \rightarrow \pi^+ \pi^-}$}

\begin{figure}
\centerline{\epsfig{figure=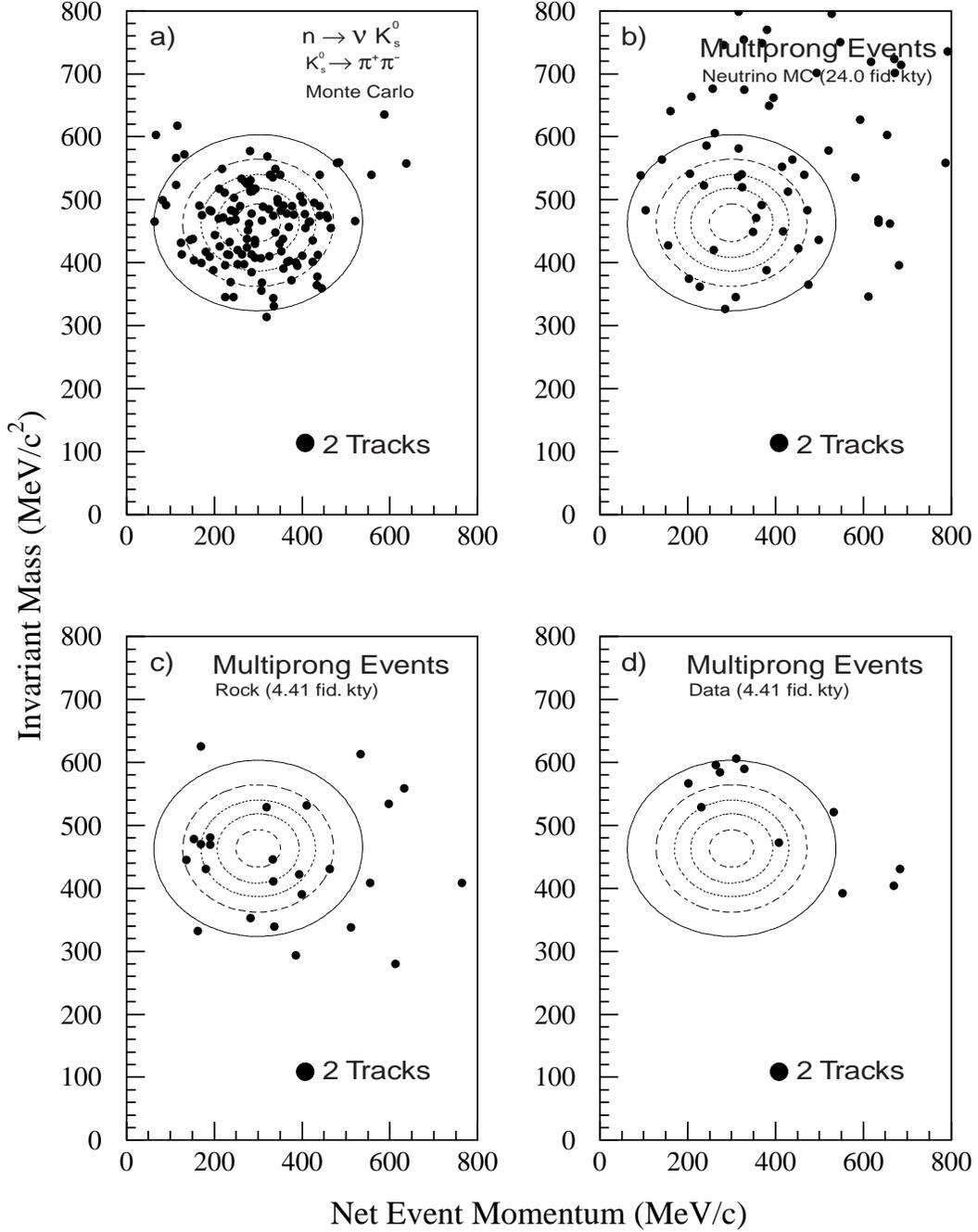,height=7in}}
\caption{For n $\rightarrow \overline{\nu}$K$^0_s$,
 K$^0_s \rightarrow \pi^+ \pi^-$, event distributions and kinematic selection
 contour in the $M_{inv}$ versus $\left| \vec{p}_{net} \right|$ plane.
 Distributions show a) the neutron decay simulation, b) atmospheric neutrino
 MC events, c) rock events, and d)the data events. Background events and data
 candidates having the two--track topology of this mode, are shown via solid
 circles.}
\label{fig:pnukspipipm}
\end{figure}

For this decay sequence,
the kinematic region in the invariant mass versus net momentum plane, which 
contains 90\% of the reconstructed neutron decays from simulation, is
delineated by the outer contour displayed in Fig.~\ref{fig:pnukspipipm}a. 
Distributions of two--track events (in which neither track is a proton) from
the atmospheric neutrino MC and from the shield--tagged rock event sample are
shown in Figs.~\ref{fig:pnukspipipm}b~and~\ref{fig:pnukspipipm}c, where
individual events are displayed as solid circles.

The survival efficiency for the $\nu$K$^0_s$ final state with
K$^0_s \rightarrow \pi^+ \pi^-$ through the
topology and kinematics requirements of Fig.~\ref{fig:pnukspipipm}a is 53\%.
Multiplying this by the K$^0_s$ branching ratio (68.6\%) and by the survival
efficiencies through triggering and
containment criteria (Table II) 
we find Soudan's overall detection efficiency 
for n $\rightarrow \nu$K$^0_s$,  K$^0_s \rightarrow \pi^+ \pi^-$ to be
$(17 \pm 2)$\%. In Fig.~\ref{fig:pnukspipipm}d we observe seven events to
satisfy the kinematic requirements of $M_{inv}$ and
$\left| \vec{p}_{net} \right|$ as defined by the
elliptical contour. Our expectation for background is 6.1 events, of which 5.1
events are predicted to have been generated by atmospheric neutrinos
(without oscillations) and 1.0 event from rock.
The distribution of data events in Fig.~\ref{fig:pnukspipipm}d is seen
to be different from one implied by the neutron decay simulation of
Fig.~\ref{fig:pnukspipipm}a. Out of seven data events which occur within
the primary contour, only two events fall within a contour which includes
80\% of the nucleon decay sample.

The mass versus momentum scatter plot of Fig.~\ref{fig:pnukspipipm}b indicates
that the neutrino--induced background for this nucleon decay mode is
significant. Of the reactions for the atmospheric neutrino MC events that
lie anywhere in Fig.~\ref{fig:pnukspipipm}b, 
we find 44\% to be single--pion or double--pion
production events (in which only one of the pions is imaged) generated by
$\nu_\mu$ or $\overline{\nu}_\mu$; an additional 42\% of these events are
misidentified $\nu_\mu$ or $\overline{\nu}_\mu$ quasi--elastics.
Electron--flavor neutrino reactions with misidentified electrons comprise
another 12\% of the background.
In the case that muon neutrinos are oscillating into other-flavor neutrinos
our MC prediction for the $\nu_{\mu}$--flavor atmospheric neutrino background is
overestimated by approximately 30\%. Thus, the above 5.1 event prediction for
the two--track topologies must be reduced to 3.6 events giving a revised rock
plus atmospheric neutrino total background estimate of 4.6 events.
The lifetime lower limit for this decay sequence with background subtraction
is, at 90\% CL, $\tau / B > 40 \times 10^{30}$~years.

\subsubsection*{$\bbox{{\rm n} \rightarrow \overline{\nu} {\rm K}^0_s}$, 
$\bbox{{\rm K}^0_s \rightarrow \pi^0 \pi^0}$}



In scanning the MC events of neutron decay which proceeds via the daughter
process K$^0_s \rightarrow \pi^0\pi^0$, we find final state four--shower and
also three--shower configurations to be the most likely topologies. In the
following we consider these two cases separately.

\begin{figure}
\centerline{\epsfig{figure=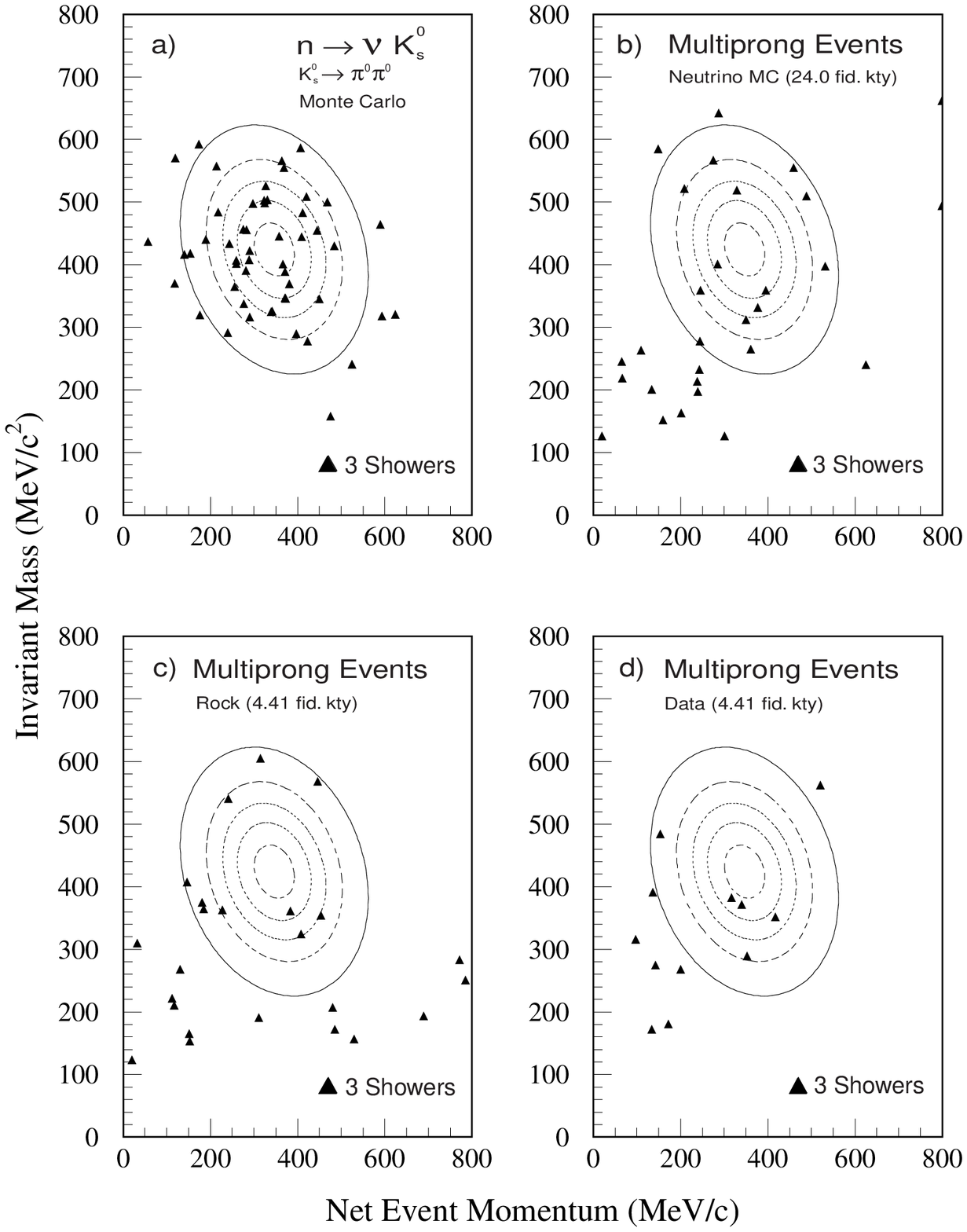,height=7in}}
\caption{For n $\rightarrow \overline{\nu}$K$^0_s$,
 K$^0 \rightarrow \pi^0 \pi^0$ yielding three visible showers. Distributions
 in the $M_{inv}$ versus $\left| \vec{p}_{net} \right|$ plane show
 three--shower events of a) the neutron decay simulation, b) the atmospheric
 neutrino MC, c) the rock events, and d) the data events.}
\label{fig:pnukspi0pi0_3s}
\end{figure}

{\it Three--shower final states:} Distributions in the invariant mass versus net
momentum plane for this case are shown in Fig.~\ref{fig:pnukspi0pi0_3s}.
Candidate events are required to have kinematics corresponding to the
occupation of the elliptical domain depicted in
Fig.~\ref{fig:pnukspi0pi0_3s}a. The detection efficiency for
n $\rightarrow  \overline{\nu}$K$^0_s$, K$^0_s \rightarrow \pi^0 \pi^0
\rightarrow$ 3 showers is 3.3\%. The number of neutrino plus rock background
events implied by Figs.~\ref{fig:pnukspi0pi0_3s}b and
\ref{fig:pnukspi0pi0_3s}c is 3.4 events. Interrogation of the atmospheric
neutrino MC events indicates that the neutrino background reactions are
generated by $\nu_e$ and $\overline{\nu}_e$ multi--pion
production charged-current events. For the cases in which charged pions were
produced, intranuclear charge exchange forced the observed final meson state
into a $\pi^0$ from which the proton decay products were reconstructed along
with the prompt electron shower. Fig.~\ref{fig:pnukspi0pi0_3s}d shows that
seven candidate events are actually observed.
From the published Kamiokande limit for this channel together with the low
detection efficiency of this experiment (Table II, row 8), 
we infer that the signal into three showers in Soudan 2 should be 0.3 events
or less. Thus, the possibility that our observed
3.6 event excess could be nucleon decay is disfavored. We set a lifetime lower
limit for this channel which is
$\tau / B > 5.6 \times 10^{30}$~years.

\begin{figure}
\centerline{\epsfig{figure=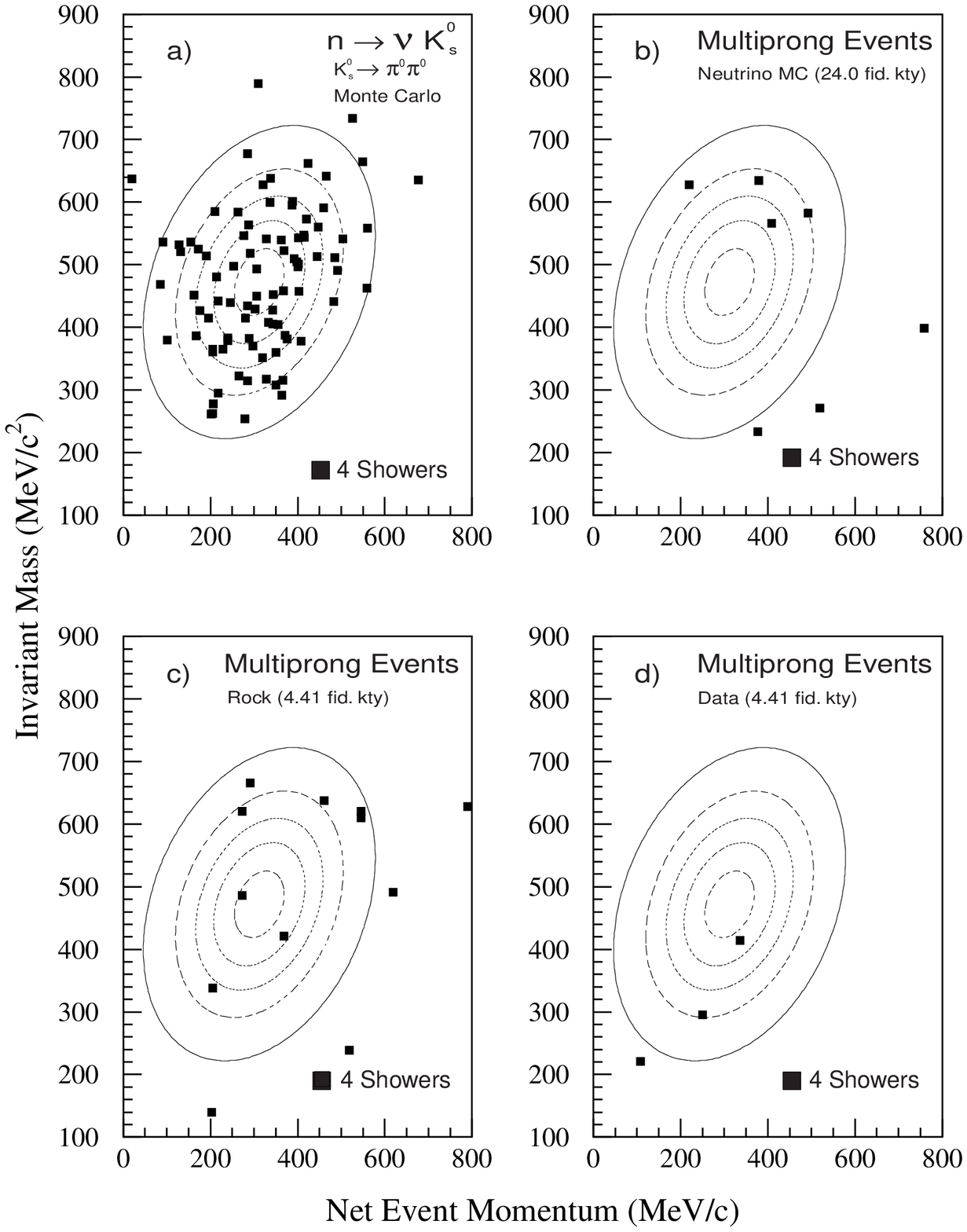,height=7in}}
\caption{For n $\rightarrow \overline{\nu}$K$^0_s$,
K$^0 \rightarrow \pi^0 \pi^0$ yielding four visible showers. Distributions in
 the $M_{inv}$ versus $\left| \vec{p}_{net} \right|$ plane show four--shower
 events of a) the neutron decay simulation, b) the atmospheric neutrino MC,
 c) the rock events, and d) the data events.}
\label{fig:pnukspi0pi0_4s}
\end{figure}

{\it Four--shower final states:} Distributions in $M_{inv}$ versus $\left|
\vec{p}_{net}  \right|$ for four-shower events of the neutron decay simulation,
of the atmospheric neutrino Monte Carlo, of the rock event sample, and of the
gold data sample, are shown in Figs.~\ref{fig:pnukspi0pi0_4s}a through
\ref{fig:pnukspi0pi0_4s}d respectively. The overall selection efficiency
for this case is 4.9\%. The number of candidate events observed is two (see
Fig.~\ref{fig:pnukspi0pi0_4s}d) whereas the total expected background is 1.1
events. For this four--shower case, the background comprising the atmospheric
neutrino Monte Carlo sample is composed of $\nu_e$ and
$\overline{\nu}_e$ multi--pion production charged-current events in which the
prompt electron shower and showers from the resulting $\pi^0$s make up the
four reconstructed prongs. Using four--shower events only, the lifetime lower
limit at 90\% CL is $\tau / B > 18 \times 10^{30}$~years.

Given the absence of a significant overall signal pattern, we have combined
all three limits from the separate cases treated above. For
n $\rightarrow \overline{\nu}$K$^0_s$ we obtain a limit of
$51 \times 10^{30}$ years. For n $\rightarrow \overline{\nu}$K$^0$ then,
the lifetime limit at 90\% CL from our
experiment is $26 \times 10^{30}$ years. Limits for
n $\rightarrow \overline{\nu} {\rm K}^0$ previously published by
 Frejus \cite{frej:numes}, Kamiokande \cite{kam:lepmes}, and
 IMB-3 \cite{imb:mcgrew} are $(15, 86, 30) \times 10^{30}$ years respectively.

\section{Statistical and systematic errors}

In the lifetime lower limit determinations of this work, there are statistical
and systematic errors which arise in event detection and final state
recognition for each decay mode, and there are errors, primarily systematic,
which arise from background estimations. We review here the variations which
may be introduced by error sources, and estimate the uncertainty
 $\Delta \tau_{N} / \tau_{N}$ on the lifetime limits obtained.

Table II displays the survival efficiencies for each of
ten distinct nucleon decay plus K decay sequences, through the selections
imposed by triggering, filtering, scanning and kinematic cuts. Knowledge of
these efficiencies is limited by the statistics and systematics
of the simulation used for
each sequence. Combined statistical errors are given with the total efficiency
of each sequence in the right--most column of Table II. 
These errors range between 11\% and 18\%; as can be seen from
Eq.~(\ref{eq:dualLimitCalc}), they contribute directly to
$\Delta \tau_{N} / \tau_{N}$. Systematic error contributions to detection
efficiencies could arise through inaccuracies in the nucleon decay simulation,
including kinematic cut inaccuracies in simulation versus data. With
regard to the latter, we note that the Monte Carlo has been tested against
data taken at the Rutherford Laboratory ISIS test beam facility
\cite{soud:garcia}. The comparisons were carried out using
beams of electrons, muons, and pions with a variety of momenta extending up to
400 MeV/$c$. For particle energies relevant to nucleon decay, the
comparisons are reassuring and would seem to exclude significant kinematic
offsets. 

Concerning nucleon decay simulation, an uncertainty arises
from inelastic intranuclear rescattering of K$^0$ within parent nuclei. A
loss of 10\% of K$^0$ born in oxygen, as a result of absorption, charge
exchange, and large-angle scattering, has been estimated by IMB \cite{imb:lK}.
Given the uncertainties in such an estimation we prefer not to include an
explicit efficiency factor, however we conclude that losses of up to 15\%
in the Soudan medium are possible.

Errors also enter the lifetime lower limit calculation
(Eqs.~(\ref{eq:dualLimitCalc})--(\ref{eq:constraints})) through the estimates of
backgrounds which include atmospheric neutrino events and the cosmic ray muon
induced rock events. Uncertainties in backgrounds have little effect on our
limits for p $\rightarrow \mu^+$K$^0_s$ and p $\rightarrow \mu^+$K$^0_l$ wherein
zero candidates are observed and low background is expected. This is not the
case for n $\rightarrow \nu$K$^0_s$ however, for which the numbers of
candidates and background events are sizable. The errors
in our rock background estimates arise primarily from the finite statistics of
shield--tagged rock control samples.
Statistical errors arise from the finite size of the atmospheric neutrino
Monte Carlo sample which corresponds to an exposure of 24.0 fiducial kiloton
years. For the individual e$^+$K$^0_s$ and $\nu$K$^0_s$ processes
(rows 3, 4 and 6-10 in Table II) 
the above-mentioned uncertainties imply
$\Delta \tau_{N} / \tau_{N} \simeq$ 15--20\%.

Background rates for neutrino reactions are based upon our atmospheric
neutrino Monte Carlo, and so the estimates are subject to uncertainties in
the absolute neutrino fluxes (20\%), in neutrino cross sections (30\%), in
the treatment of intranuclear rescattering (30\%), and in the treatment of
other nuclear effects, e.g. Fermi motion and Pauli blocking (20\%)
\cite{hugh:thesis}. Inherent to our correction of
 $(\nu_\mu + \overline{\nu}_\mu)$ charged-current background rates to account
for neutrino oscillations, is the uncertainty (20\%) in the Soudan 2
flavor ratio. Uncertainties in the backgrounds, although sizable,
affect the lifetime lower
limit calculation through variation of the $b_i$ of Eq.~(\ref{eq:mucalc});
the variation propagates to the $\mu_i$ of Eq.~(\ref{eq:constraints}) via
solution of Eq.~(\ref{eq:mucalc}) with constraint
(\ref{eq:constraints}) -- a non--linear relation.
Propagating the above listed uncertainties through for the
n $\rightarrow \nu$K$^0_s$ case, we find
$\Delta \tau_{N} / \tau_{N} \simeq 16\%$.

We conclude that the uncertainty $\Delta \tau_{N} / \tau_{N}$ on the lifetime
lower limits reported here arising from the sources described above
is approximately 20\% for proton decay into
$\mu^+$K$^0_s$ and $\mu^+$K$^0_l$, and may be as large as 32\% for
e$^+$K$^0_s$ and e$^+$K$^0_l$ final states, and for neutron decay to
$\nu$K$^0_s$.

\section{Discussion and Summary}

      We report results from a search for nucleon decay into two-body, 
lepton-plus-K$^0$ final states.   Current SUSY Grand Unified Theories propose
that these nucleon decay modes occur and may be predominant
(Refs. 7--12); 
our investigation probes the lower range of published lifetime predictions.
The search is carried out using Soudan 2's fine-grained tracking 
calorimeter of honeycomb lattice geometry, a detector which images 
non-relativistic as well as relativistic charged tracks and provides
$dE/dx$ ionization sampling.  The fiducial exposure utilized, 4.41 kiloton-
years, is twice as large as any previously reported using the 
tracking calorimeter technique.  For each of the decay channels 
investigated, selections have been designed which reduce the neutrino and 
cosmic ray background to levels of few events, while maintaining sufficient
detection efficiency to allow a sensitive search.   In each of the channels we
find zero or small numbers of nucleon decay candidates; the latter occur
at or below rates calculated for background.   No evidence for a nucleon
decay signal is observed, and we report lifetime lower limits $\tau/B$ at
90\% CL as summarized in Table V. 
In the Table, the limit
$\tau/B$ given for each K$^0_s$ channel is larger than the limit for the
corresponding K$^0$ channel due to the branching fraction
$B({\rm  K}^0 \rightarrow {\rm K}^0_s)$ = 0.5\,. For the K$^0$ channels,
the listed $\tau/B$ values represent our best measured limits obtained by
summing the K$^0_s$ and K$^0_l$ channels. Under the assumption of CP
invariance in the decay, i.e.
$B({\rm K}^0 \rightarrow {\rm K}^0_s) = B({\rm K}^0 \rightarrow {\rm K}^0_l) =
0.5$, better limits equal to the K$^0_s$ limits could be inferred.

For the p $\rightarrow \mu^+$K$^0_s$ and p $\rightarrow \mu^+$K$^0_l$ channels
we estimate the background to be less than one event and we find no candidate
events. The lifetime lower limits at 90\% CL thus obtained for
($\mu^+$K$^0_s$, $ \mu^+$K$^0_l$) are $(150, 83) \times 10^{30}$ years.
Previous limits for these channels, reported by Frejus \cite{frej:chlep} and
listed by the Particle Data Group (PDG)\cite{pdg:rpp}, are
$(64, 44) \times 10^{30}$ years. By combining these channel limits, the limit
for p $\rightarrow \mu^+$K$^0$ in the Soudan 2 experiment is obtained:
$\tau/B > 120 \times 10^{30}$ years. Previous limits for this mode from
Kamiokande, Frejus, and IMB-3 are $(120, 54, 120) \times 10^{30}$ years
respectively \cite{kam:lepmes,frej:chlep,imb:mcgrew}. 

For p $\rightarrow$ e$^+$K$^0_s$ and p $\rightarrow$ e$^+$K$^0_l$ channels
we estimate background rates of one event and 3.5 events, and we observe
one candidate and two candidates respectively. Our lifetime lower limits for
(e$^+$K$^0_s$, e$^+$K$^0_l$) are $(120, 51) \times 10^{30}$ years, an
improvement upon the limits $(76, 44) \times 10^{30}$ years
established previously by Frejus \cite{frej:chlep,pdg:rpp}. For the
mode p $\rightarrow$ e$^+$K$^0$, we obtain
$\tau/B > 85 \times 10^{30}$ years, to be compared with previous limits
by Kamiokande, Frejus, and IMB-3: $(150, 60, 31) \times 10^{30}$
years respectively \cite{kam:lepmes,frej:chlep,imb:mcgrew}. 

For neutron decay into $\nu$K$^0_s$ we observe candidates appearing at rates
which are compatible with background estimates.
Our lifetime lower limit for $\nu$K$^0$ is 
$\tau/B > 26 \times 10^{30}$ years. Although the Soudan 2 limit for this mode
is the most restrictive ever obtained using the tracking calorimeter technique,
limits of $(86, 30) \times 10^{30}$ years have been reported by
Kamiokande and IMB-3 \cite{kam:lepmes,imb:mcgrew}.


\section*{Acknowledgements}

This work was supported by the U.S Department of Energy, the U.K. Particle
 Physics and Astronomy Research Council, and the State and University of
 Minnesota. We also wish to thank the Minnesota Department of Natural
 Resources for allowing us to use the facilities of the Soudan Underground
Mine State Park.

\clearpage




\begin{table}[ht]
\begin{center}

\footnotesize
\tabcolsep 0.01cm

\begin{tabular}{cccc}
\\  
Model & Leading Mode(s) & Predicted Lifetimes & Authors [Ref.] \\
\\
\tableline 
\\
SUSY $SU(5)$       & n $\rightarrow \overline{\nu}$K$^0$,
p $\rightarrow \overline{\nu}$K$^+$ & $\sim 10^{31}$y & 
Hisano, Murayama (1993) \cite{theory:Hisano_and_Murayama}\\
\\
Discrete $(S_3)^{3}$ & p $\rightarrow$ e$^+$K$^0$ & $\sim 10^{32}$y &
Carone {\it et al.} (1996) \cite{susy:carone}\\
\\
SUSY $SO(10)$ with       & p $\rightarrow \ell^+$K$^0$ & -- &
Babu \& Barr (1996) \cite{theory:babu_barr} \\
``realistic masses'' & & & \\
\\
SUSY $SO(10)$       & n $\rightarrow \nu$K$^0$, p $\rightarrow \nu$K$^+$
& $\sim 10^{32}$y & Lucas \& Rabi (1997)
\cite{theory:lucas_and_rabi}\\
\\
Anomalous $U(1)$    & p $\rightarrow \mu^+$K$^0$ & $<10^{32}$y & Irges {\it et
al.} (1998) \cite{theory:Irges}\\
from superstrings & & & \\
\\
SUSY $SO(10)$ with       & p $\rightarrow \ell^+$K$^0$, $\ell^+ \pi^0$,
  $\ell^+ \eta$ & -- & Babu {\it et al.} (1998) \cite{babu_pati} \\
see-saw $\nu$ masses & & & \\
\end{tabular}
\caption{SUSY model predictions for nucleon decay into `lepton + K$^0$'
 final states.}
\end{center}
\label{tbl:table1}
\end{table}


\begin{table}[ht]
\begin{center}
\footnotesize
\tabcolsep 0.10cm
\footnotesize

\begin{tabular}{lcccccccc}
\\  
 Decay&Daughter&Hardware &Contain-&Event &Topology  &
 Kinematic&BR&\ $\epsilon\times$BR \\
  Mode&Process & Trigger & ment &Quality &Selection & Cuts     &  &  \\
& & & Filter & Scans & & & & \\
\\
\tableline
\\
 p $\rightarrow \mu^+$K$^0_s$ & K$^0_s \rightarrow \pi^+\pi^-$
 & 0.97 & 0.67 & 0.72 & 0.56 & 0.88 & 0.69 & 0.16$\pm$0.03 \\
 p $\rightarrow \mu^+$K$^0_s$ & K$^0_s \rightarrow \pi^0\pi^0$
 & 0.99 & 0.57 & 0.67 & 0.63 & 0.80 & 0.31 & 0.06$\pm$0.01\\
 p $\rightarrow$ e$^+$K$^0_s$ & K$^0_s \rightarrow \pi^+\pi^-$
 & 0.97 & 0.68 & 0.73 & 0.51 & 0.86 & 0.69 & 0.15$\pm$0.03\\
 p $\rightarrow$ e$^+$K$^0_s$ & K$^0_s \rightarrow \pi^0\pi^0$
 & 0.99 & 0.61 & 0.70 & 0.78 & 0.75 & 0.31 & 0.08$\pm$0.01\\
 p $\rightarrow \mu^+$K$^0_l$ & K$^0_l \rightarrow$ int.
 & 0.99 & 0.59 & 0.77 & 0.53 & 0.53 & 1.0 & 0.12$\pm$0.01\\
 p $\rightarrow$ e$^+$K$^0_l$ & K$^0_l \rightarrow$ int.
 & 0.97 & 0.61 & 0.74 & 0.48 & 0.50 & 1.0 & 0.11$\pm$0.01\\
 n $\rightarrow \overline{\nu}$K$^0_s$ & K$^0_s \rightarrow \pi^+\pi^-$
 & 0.87 & 0.73 & 0.72 & 0.56 & 0.94 & 0.69 & 0.17$\pm$0.02\\
 n $\rightarrow \overline{\nu}$ K$^0_s$ &  K$^0_s \rightarrow \pi^0\pi^0$ (3 
$\gamma$)
 & 0.93 & 0.68 & 0.61 & 0.31 & 0.90 & 0.31 & 0.03$\pm$0.01\\
 n $\rightarrow \overline{\nu}$ K$^0_s$ &  K$^0_s \rightarrow \pi^0\pi^0$ (4
$\gamma$)
 & 0.93 & 0.68 & 0.61 & 0.44 & 0.94 & 0.31 & 0.05$\pm$0.01\\
\end{tabular}
\caption{Event sample survival fractions (columns 3 through 7) and channel
detection efficiencies (column 9) for nucleon decay final states
containing K$^0$ mesons. The events simulated for these Monte Carlo
samples were generated within the full volume of the detector, consequently
the survival rates reported in the Containment and Event Quality columns
include removal of events which extend outside the fiducial volume.}
\end{center}
\label{tbl:table2}
\end{table}


\begin{table}
\begin{center}
\begin{tabular}{ccccc}
\\
     Event   & p $\rightarrow \mu^+$K$_l$ & Atm. $\nu$  & Rock   & Data   \\
Selections &         Simulation            & Background  & Background & Events
\\
& (493 evts) & (4.41 fid.kty) & (4.41 fid.kty) \\
\\
\tableline 
\\
Track--Gap--Interact. Topology                 & 110 & 2.2 & 0.8 &  1  \\
$ 35 \leq l_{\mu^+} \leq 120$ cm               & 105 & 2.0 & 0.6 &  1  \\
15 cm$ \leq$ K$^0_l$ path gap                  &  80 & 0.9  & 0.5  &  1  \\
$ 200 \leq$ K$^0_l$ ``energy'' $\leq 1400$ MeV &  69 & 0.4  & 0.4  &  0  \\
$ {\rm cos}({\rm K}^0_l \cdot {\textstyle interaction})
 \geq 0$ &  63 & 0.4  & 0.3  &  0  \\
$ {\rm cos}({\rm K}^0_l \cdot \mu^+) \leq 0$   &  58 & 0.2  & 0.2 &  0  \\
\end{tabular}
\caption{For the p $\rightarrow \mu^+$K$^0_l$ search, the evolution of
event sample populations resulting from successive application of
topology and kinematic selections. The four columns to the right of each
Event Selection show survival populations for the nucleon decay simulation,
for neutrino background events of the atmospheric $\nu$ Monte Carlo,
for the rock background (estimated from shield-tagged data), and for gold
data events. 
The atmospheric neutrino and rock background populations are normalized
to the 4.41 fiducial kiloton-year exposure.}
\end{center}
\label{tbl:table3}
\end{table}


\begin{table}
\begin{center}
\begin{tabular}{ccccc}
\\
Event        & p $\rightarrow$ e$^+$ K$^0_l$ & Atm. $\nu$ &  Rock  & Data  \\
Selections   &       Simulation            & Background & Background & Events \\
& (493 evts) & (4.41 fid.kty) & (4.41 fid.kty) \\
\\
\tableline 
\\
Shower--Gap--Interact. Topology                & 101 & 17.5 & 3.2 & 25  \\
$ 120 \leq P_{{\rm e}^+} \leq 500$ MeV/$c$     &  92 & 17.0 & 3.1 & 25  \\
 15 $cm \leq$ K$^0_l$ path gap                 &  67 & 14.0  & 2.7 & 18  \\
$ 200 \leq$ K$^0_l$ ``energy'' $\leq 1100$ MeV &  61 & 6.7  & 1.8 &  8  \\
$ {\rm cos}({\rm K}^0_l \cdot {\textstyle interaction})
 \geq 0$  &  55 & 2.9  & 0.9 &  3  \\
$ {\rm cos}({\rm K}^0_l \cdot {\rm e}^+) \leq 0$
          &  51 & 2.6  & 0.8 &  2  \\
\end{tabular}
 
\caption{For the p $\rightarrow$ e$^+$K$^0_l$ search,
 the evolution of event sample populations resulting from successive
 application of topology and kinematic selections.
Rates for surviving atmospheric neutrino and cosmic-ray neutron-induced
backgrounds are for the 4.41 fiducial kiloton-year exposure.
The net backgrounds (bottom row, columns two and three) are higher than
those obtained for $\mu^+$K$^0_l$ (see Table III).}
\end{center}
\label{tbl:table4}
\end{table}


\begin{table}
\begin{center}
\begin{tabular}{lccccccc}
\\ 
Decay Mode&Final State&$\epsilon \times$B.R.&$\nu_{\textstyle Bk}$&
Total$_{\textstyle Bk}$&Data&$\tau /B \times 10^{30}$y\\
\\ 
\tableline
\\ 
p $\rightarrow \mu^+$K$_s^0$ & $\mu^+\pi^+\pi^-$ & 0.16 & $<0.2$  & $<0.2$  &
0 & 150 \\
& $\mu^+\pi^0\pi^0$ & 0.06 & 0.6     &  0.6 & 0 &    \\ 
\\ 
p $\rightarrow$ e$^+$ K$_s^0$  
 & e$^+\pi^+\pi^-$  & 0.15 & 0.6 & 0.7 & 1 & 120 \\ 
 & e$^+\pi^0\pi^0$  & 0.08 & 0.4 & 0.6 & 0 &     \\ 
\\ 
p $\rightarrow \mu^+$K$_l^0$ & K$^0_l \rightarrow$ interaction
& 0.12 & 0.2 & 0.4 & 0 & 83  \\ 
\\ 
p $\rightarrow$ e$^+$K$_l^0$   & K$^0_l \rightarrow$ interaction
& 0.11 & 2.6 & 3.5 & 2 & 51 \\ 
\\ 
p $\rightarrow \mu^+$K$^0$   & $\mu^+({\rm K}^0_s + {\rm K}^0_l)$ & 0.17
& $<$0.9 & $<$1.2 & 0 & 120 \\
p $\rightarrow$ e$^+$K$^0$   & {\rm e}$^+({\rm K}^0_s + {\rm K}^0_l)$ & 0.17
&3.5 & 4.9 & 3 & 85 \\
\\ 
n $\rightarrow \nu$K$_s^0$  & $\pi^+\pi^-$ & 0.17 & 3.6(5.1) & 4.6(6.1)
 & 7 & 51(59) \\
 (three -- showers)         & $\pi^0\pi^0$ & 0.03 & 2.6 & 3.4 & 7 &    \\
 (four -- showers)          & $\pi^0\pi^0$ & 0.05 & 0.6 &  1.1 & 2 &  \\
n $\rightarrow \nu$K$^0$ & $\nu({\rm K}^0_s + {\rm K}^0_l)$ 
& 0.13 & 6.8(8.3) & 9.1(10.6) & 16 & 26(29) \\
\end{tabular}

\caption{Background--subtracted lifetime lower limits at 90\%
 confidence level from Soudan 2.
Correction of neutrino background for $\nu_\mu$-flavor depletion by oscillations
has significant effect for n $\rightarrow \nu{\rm K}^0_s$; values without this
correction are given in parentheses. For K$^0$ channels, we list our
best measured limits obtained by summing the K$^0_s$ and K$^0_l$ channels.}
\end{center}
\label{tbl:table5}
\end{table}


\end{document}